\documentclass[12pt,thmsa]{article}
\usepackage{amssymb}

\newtheorem{theorem}{Theorem}
\newcommand{\proof}{{\noindent \bf{Proof. }}}

\begin{document}

\title{An irreducible approach to second-order reducible second-class
constraints}
\author{C. Bizdadea\thanks{%
E-mail address: bizdadea@central.ucv.ro}, E. M. Cioroianu\thanks{%
E-mail address: manache@central.ucv.ro}, S. O. Saliu\thanks{%
E-mail address: osaliu@central.ucv.ro}, \and S. C. S\u{a}raru\thanks{%
E-mail address: scsararu@central.ucv.ro}, O. B\u{a}lu\c{s}\\
Faculty of Physics, University of Craiova\\
13 Al. I. Cuza Str., Craiova 200585, Romania}
\date{}
\maketitle

\begin{abstract}
An irreducible canonical approach to second-order reducible
second-class constraints is given. The procedure is exemplified on
gauge-fixed three-forms.
\end{abstract}

\section{Introduction}

The canonical approach to systems with reducible second-class
constraints is quite intricate, demanding a modification of the
usual rules as the matrix of the Poisson brackets among the
constraints is not invertible. Thus, it is necessary to isolate a
set of independent constraints and then construct the Dirac bracket
\cite{1,2} with respect to this set. The split of the constraints
may lead to the loss of important symmetries, so it should be
avoided. As shown in \cite{3,4,5,6,7,9}, it is however possible to
construct the Dirac bracket in terms of a noninvertible matrix
without separating the independent constraint functions. A third
possibility is to substitute the reducible second-class constraints
by some irreducible ones and further work with the Dirac bracket
based on the irreducible constraints. This idea, suggested in
\cite{8} mainly in the context of 2- and 3-form gauge fields, has
been developed in a general manner only for first-order reducible
second-class constraints \cite{EPL}.

In this paper, we give an irreducible approach to second-order
reducible second-class constraints. Our strategy includes three main
steps. First, we express the Dirac bracket for the reducible system
in terms of an invertible matrix. Second, we construct an
intermediate second-order reducible second-class system on a larger
phase space and establish the equality between the original Dirac
bracket and that corresponding to the intermediate theory. Third, we
prove that there exists an irreducible second-class constraint set
equivalent to the intermediate one, such that the corresponding
Dirac brackets coincide. These three steps enforce the fact that the
fundamental Dirac brackets derived within the irreducible and
original reducible settings coincide.

The present paper is organized into five sections. In Section
\ref{rev}, we briefly review the procedure for first-order reducible
second-class constraints. Section \ref{secord} is the `hard core' of
the paper. Here, we approach second-order reducible second-class
constraints by implementing the three main steps mentioned above. In
Section \ref{exam}, we exemplify in detail the general procedure
from Section \ref{secord} in the case of gauge-fixed three-form
gauge fields. Section \ref{conc} ends the paper with the main
conclusions.

\section{First-order reducible second-class constraints: a brief review
\label{rev}}

\subsection{Dirac bracket for first-order reducible second-class constraints}

We start with a system locally described by $N$ canonical pairs $%
z^{a}=\left( q^{i},p_{i}\right) $, subject to some constraints
\begin{equation}
\chi _{\alpha _{0}}\left( z^{a}\right) \approx 0,\;\alpha _{0}=1,\ldots
,M_{0}.  \label{1}
\end{equation}%
For simplicity, we take all the phase-space variables to be bosonic.
However, our analysis can be extended to fermionic degrees of freedom modulo
including some appropriate phase factors. We choose the scenario of systems
with a finite number of degrees of freedom only for notational simplicity,
but our approach is equally valid for field theories. In addition, we
presume that the functions $\chi _{\alpha _{0}}$ are not all independent,
but there exist some nonvanishing functions $Z_{\;\;\alpha _{1}}^{\alpha
_{0}}$ such that
\begin{equation}
Z_{\;\;\alpha _{1}}^{\alpha _{0}}\chi _{\alpha _{0}}=0,\;\alpha
_{1}=1,\ldots ,M_{1}.  \label{2}
\end{equation}%
Moreover, we assume that $Z_{\;\;\alpha _{1}}^{\alpha _{0}}$ are all
independent and (\ref{2}) are the only reducibility relations with
respect to the constraints (\ref{1}). These constraints are purely
second class if any maximal, independent set of $M_{0}-M_{1}$
constraint functions $\chi _{A} $ ($A=1,\ldots ,M_{0}-M_{1}$) among
the $\chi _{\alpha _{0}}$ is such that the matrix
\begin{equation}
C_{AB}^{\left( 1\right) }=\left[ \chi _{A},\chi _{B}\right] ,  \label{3}
\end{equation}%
is invertible. Here and in the following the symbol $\left[ ,\right]
$ denotes the Poisson bracket. In terms of independent constraints,
the Dirac bracket takes the form
\begin{equation}
\left[ F,G\right] ^{\left( 1\right) \ast }=\left[ F,G\right] -\left[ F,\chi
_{A}\right] M^{\left( 1\right) AB}\left[ \chi _{B},G\right] ,  \label{4}
\end{equation}%
where $M^{\left( 1\right) AB}C_{BC}^{\left( 1\right) }\approx \delta
_{\;\;C}^{A}$. In the previous relations we introduced an extra index, $%
\left( 1\right) $, having the role to emphasize that the Dirac bracket (\ref%
{4}) is based on a first-order reducible second-class constraint set. We can
rewrite the Dirac bracket (\ref{4}) without finding a definite subset of
independent second-class constraints as follows. We start with the matrix
\begin{equation}
C_{\alpha _{0}\beta _{0}}^{\left( 1\right) }=\left[ \chi _{\alpha _{0}},\chi
_{\beta _{0}}\right] ,  \label{5}
\end{equation}%
which clearly is not invertible because
\begin{equation}
Z_{\;\;\alpha _{1}}^{\alpha _{0}}C_{\alpha _{0}\beta _{0}}^{\left( 1\right)
}\approx 0.  \label{6}
\end{equation}%
If $\bar{a}_{\alpha _{0}}^{\;\;\alpha _{1}}$ is a solution to the equation
\begin{equation}
\bar{a}_{\alpha _{0}}^{\;\;\alpha _{1}}Z_{\;\;\beta _{1}}^{\alpha
_{0}}\approx \delta _{\;\;\beta _{1}}^{\alpha _{1}},  \label{7}
\end{equation}%
then we can introduce a matrix \cite{6} $M^{\left( 1\right) \alpha _{0}\beta
_{0}}$ through the relation
\begin{equation}
M^{\left( 1\right) \alpha _{0}\beta _{0}}C_{\beta _{0}\gamma _{0}}^{\left(
1\right) }\approx \delta _{\;\;\gamma _{0}}^{\alpha _{0}}-Z_{\;\;\alpha
_{1}}^{\alpha _{0}}\bar{a}_{\gamma _{0}}^{\;\;\alpha _{1}}\equiv
d_{\;\;\gamma _{0}}^{\alpha _{0}},  \label{8}
\end{equation}%
with $M^{\left( 1\right) \alpha _{0}\beta _{0}}=-M^{\left( 1\right) \beta
_{0}\alpha _{0}}$. Then, formula \cite{6}
\begin{equation}
\left[ F,G\right] ^{\left( 1\right) \ast }=\left[ F,G\right] -\left[ F,\chi
_{\alpha _{0}}\right] M^{\left( 1\right) \alpha _{0}\beta _{0}}\left[ \chi
_{\beta _{0}},G\right] ,  \label{10}
\end{equation}%
defines the same Dirac bracket like (\ref{4}) on the surface (\ref{1}). We
remark that there exist some ambiguities in defining the matrix $M^{\left(
1\right) \alpha _{0}\beta _{0}}$ since if we make the transformation
\begin{equation}
M^{\left( 1\right) \alpha _{0}\beta _{0}}\rightarrow M^{\left( 1\right)
\alpha _{0}\beta _{0}}+Z_{\;\;\alpha _{1}}^{\alpha _{0}}q^{\alpha _{1}\beta
_{1}}Z_{\;\;\beta _{1}}^{\beta _{0}},  \label{10wq}
\end{equation}%
with $q^{\alpha _{1}\beta _{1}}$ some completely antisymmetric functions,
then equation (\ref{8}) is still satisfied.

At this stage it is useful to make some comments. First, we remark that
relations (\ref{7}) and (\ref{8}) yield
\begin{equation}
M^{\left( 1\right) \alpha _{0}\beta _{0}}C_{\beta _{0}\gamma _{0}}^{\left(
1\right) }Z_{\;\;\beta _{1}}^{\gamma _{0}}\approx 0,  \label{13}
\end{equation}%
which ensures the fact that the rank of $M^{\left( 1\right) \alpha _{0}\beta
_{0}}C_{\beta _{0}\gamma _{0}}^{\left( 1\right) }$ is equal to the number of
independent second-class constraints, i.e.,
\begin{equation}
\mathrm{rank}\left( M^{\left( 1\right) \alpha _{0}\beta _{0}}C_{\beta
_{0}\gamma _{0}}^{\left( 1\right) }\right) \approx M_{0}-M_{1}.  \label{10a}
\end{equation}%
Second, by means of (\ref{8}) we deduce the relation
\begin{equation}
\left[ \chi _{\alpha _{0}},G\right] ^{\left( 1\right) \ast }\approx -\bar{a}%
_{\alpha _{0}}^{\;\;\alpha _{1}}\left[ Z_{\;\;\alpha _{1}}^{\beta _{0}},G%
\right] \chi _{\beta _{0}},  \label{11}
\end{equation}%
which ensures
\begin{equation}
\left[ \chi _{\alpha _{0}},G\right] ^{\left( 1\right) \ast }=0,\;\mathrm{%
for\;any}\;G,  \label{11a}
\end{equation}%
on the second-class surface, as required by the general properties of the
Dirac bracket. Third, we remark that, in spite of the fact that the matrix $%
C_{\alpha _{0}\beta _{0}}^{\left( 1\right) }$ is not invertible, the Dirac
bracket expressed by (\ref{10}) still satisfies Jacobi's identity%
\begin{equation}
\left[ \left[ F,G\right] ^{\left( 1\right) \ast },P\right] ^{\left( 1\right)
\ast }+\left[ \left[ P,F\right] ^{\left( 1\right) \ast },G\right] ^{\left(
1\right) \ast }+\left[ \left[ G,P\right] ^{\left( 1\right) \ast },F\right]
^{\left( 1\right) \ast }\approx 0  \label{c1}
\end{equation}%
on surface (\ref{1}). The proof follows the same line like in the
irreducible case. Let%
\begin{equation}
\bar{F}=F+u^{\beta _{0}}\chi _{\beta _{0}},  \label{c2}
\end{equation}%
be a function such that%
\begin{equation}
\left[ \bar{F},\chi _{\alpha _{0}}\right] \approx 0.  \label{c3}
\end{equation}%
Thus, in order to construct $\bar{F}$ we must solve the equation
\begin{equation}
u^{\beta _{0}}C_{\beta _{0}\alpha _{0}}^{\left( 1\right) }\approx -\left[
F,\chi _{\alpha _{0}}\right] .  \label{c4}
\end{equation}%
Based on%
\begin{equation}
d_{\;\;\alpha _{0}}^{\lambda _{0}}\chi _{\lambda _{0}}=\chi _{\alpha _{0}},
\label{c5}
\end{equation}%
it follows in a simple manner that the solution to equation
(\ref{c4}) is
given by%
\begin{equation}
u^{\beta _{0}}=\left[ F,\chi _{\lambda _{0}}\right] M^{\left( 1\right) \beta
_{0}\lambda _{0}},  \label{c6}
\end{equation}%
which further leads to%
\begin{equation}
\bar{F}=F+\left[ F,\chi _{\beta _{0}}\right] M^{\left( 1\right) \alpha
_{0}\beta _{0}}\chi _{\alpha _{0}}.  \label{c7}
\end{equation}%
Relying on (\ref{c5}) and (\ref{c7}), by direct computation we arrive at the
relation%
\begin{equation}
\left[ \left[ F,G\right] ^{\left( 1\right) \ast },P\right] ^{\left( 1\right)
\ast }\approx \left[ \left[ \bar{F},\bar{G}\right] ,\bar{P}\right] ,
\label{c8}
\end{equation}%
which indicates that identity (\ref{c1}) is ensured by Jacobi's
identity corresponding to the Poisson bracket for the functions
$\bar{F}$, $\bar{G}$ and $\bar{P}$. We mention that the key point of
the proof of Jacobi's identity (\ref{c1}) is represented by relation
(\ref{c5}).

\subsection{Irreducible analysis of first-order reducible second-class
constraints\label{ford}}

First-order reducible second-class constraints can be approached in an
irreducible manner, as it has been shown in \cite{EPL}. To this end, one
starts from the solution to equation (\ref{7})
\begin{equation}
\bar{a}_{\alpha _{0}}^{\;\;\alpha _{1}}=\bar{D}_{\;\;\gamma _{1}}^{\alpha
_{1}}a_{\alpha _{0}}^{\;\;\gamma _{1}},  \label{11m}
\end{equation}%
where $a_{\alpha _{0}}^{\;\;\gamma _{1}}$ are some functions chosen such
that
\begin{equation}
\mathrm{rank}\left( Z_{\;\;\alpha _{1}}^{\alpha _{0}}a_{\alpha
_{0}}^{\;\;\gamma _{1}}\right) =M_{1}  \label{12}
\end{equation}%
and $\bar{D}_{\;\;\gamma _{1}}^{\beta _{1}}$ stands for the inverse of $%
Z_{\;\;\alpha _{1}}^{\alpha _{0}}a_{\alpha _{0}}^{\;\;\gamma _{1}}$.
In order to develop an irreducible approach it is necessary to
enlarge the original phase space with some new variables $\left(
Y_{\alpha _{1}}\right) _{\alpha _{1}=1,\ldots ,M_{1}}$, endowed with
the Poisson brackets
\begin{equation}
\left[ Y_{\alpha _{1}},Y_{\beta _{1}}\right] =\Gamma _{\alpha _{1}\beta
_{1}},  \label{12xx}
\end{equation}%
where $\Gamma _{\alpha _{1}\beta _{1}}$ are the elements of an invertible,
antisymmetric matrix that may depend on the newly added variables.
Consequently, one constructs the constraints
\begin{equation}
\bar{\chi}_{\alpha _{0}}=\chi _{\alpha _{0}}+a_{\alpha _{0}}^{\;\;\alpha
_{1}}Y_{\alpha _{1}}\approx 0,  \label{35}
\end{equation}%
which are second-class and, essentially, irreducible. Following the line
exposed in \cite{EPL} it can be shown that the Dirac bracket associated with
the irreducible constraints takes the form
\begin{equation}
\left. \left[ F,G\right] ^{\left( 1\right) \ast }\right\vert _{\mathrm{ired}%
}=\left[ F,G\right] -\left[ F,\bar{\chi}_{\alpha _{0}}\right] \mu
^{\left( 1\right) \alpha _{0}\beta _{0}}\left[ \bar{\chi}_{\beta
_{0}},G\right] ,\label{31}
\end{equation}%
and it is (weakly) equal to the original Dirac bracket (\ref{4}),
\begin{equation}
\left[ F,G\right] ^{\left( 1\right) \ast }\approx \left. \left[ F,G\right]
^{\left( 1\right) \ast }\right\vert _{\mathrm{ired}}.  \label{31zx}
\end{equation}%
In (\ref{31}) the quantities $\mu ^{\left( 1\right) \alpha _{0}\beta _{0}}$
are the elements of an invertible, antisymmetric matrix, expressed by
\begin{equation}
\mu ^{\left( 1\right) \alpha _{0}\beta _{0}}\approx M^{\left( 1\right)
\alpha _{0}\beta _{0}}+Z_{\;\;\lambda _{1}}^{\alpha _{0}}\bar{D}_{\;\;\beta
_{1}}^{\lambda _{1}}\Gamma ^{\beta _{1}\gamma _{1}}\bar{D}_{\;\;\gamma
_{1}}^{\sigma _{1}}Z_{\;\;\sigma _{1}}^{\beta _{0}},  \label{31qa}
\end{equation}%
with $\Gamma ^{\beta _{1}\gamma _{1}}$ the inverse of $\Gamma _{\alpha
_{1}\beta _{1}}$. Formula (\ref{31zx}) is essential in our context because
it proves that one can indeed approach first-order reducible second-class
constraints in an irreducible fashion.

\section{Second-order reducible second-class constraints\label{secord}}

\subsection{Reducible approach}

\subsubsection{Dirac bracket for second-order reducible second-class
constraints}

In the following we will generalize the previous approach to the
case of second-order reducible second-class constraints. This means
that not all of the first-order reducibility functions
$Z_{\;\;\alpha _{1}}^{\alpha _{0}}$ are independent. Beside the
first-order reducibility relations (\ref{2}), there appear also the
second-order reducibility relations
\begin{equation}
Z_{\;\;\alpha _{2}}^{\alpha _{1}}Z_{\;\;\alpha _{1}}^{\alpha _{0}}\approx
0,\;\alpha _{2}=1,\ldots ,M_{2}.  \label{11x}
\end{equation}%
We will assume that the reducibility stops at order 2, so the functions $%
Z_{\;\;\alpha _{2}}^{\alpha _{1}}$ are by hypothesis taken to be
independent. It is understood that $Z_{\;\;\alpha _{2}}^{\alpha
_{1}}$'s define a complete set of reducibility functions for
$Z_{\;\;\alpha _{1}}^{\alpha _{0}}$. In this situation, the number
of independent second-class constraints is equal to
$M_{0}-M_{1}+M_{2}$. As a consequence,
we can work with a Dirac bracket of the type (\ref{4}), but in terms of $%
M_{0}-M_{1}+M_{2}$ independent functions $\chi _{A}$
\begin{equation}
\left[ F,G\right] ^{\left( 2\right) \ast }=\left[ F,G\right] -\left[ F,\chi
_{A}\right] M^{\left( 2\right) AB}\left[ \chi _{B},G\right] ,\;A=1,\ldots
,M_{0}-M_{1}+M_{2},  \label{11b}
\end{equation}%
where $M^{\left( 2\right) AB}C_{BC}^{\left( 2\right) }\approx \delta
_{\;\;C}^{A}$, with $C_{AB}^{\left( 2\right) }=\left[ \chi _{A},\chi _{B}%
\right] $. It is obvious that the matrix
\begin{equation}
C_{\alpha _{0}\beta _{0}}^{\left( 2\right) }=\left[ \chi _{\alpha _{0}},\chi
_{\beta _{0}}\right]  \label{11d}
\end{equation}%
satisfies the relations
\begin{equation}
Z_{\;\;\alpha _{1}}^{\alpha _{0}}C_{\alpha _{0}\beta _{0}}^{\left( 2\right)
}\approx 0,  \label{11e}
\end{equation}%
so its rank is equal to $M_{0}-M_{1}+M_{2}$.

Let $\bar{A}_{\alpha _{1}}^{\;\;\alpha _{2}}$ be a solution of the equation
\begin{equation}
Z_{\;\;\beta _{2}}^{\alpha _{1}}\bar{A}_{\alpha _{1}}^{\;\;\alpha
_{2}}\approx \delta _{\;\;\beta _{2}}^{\alpha _{2}}  \label{a2}
\end{equation}%
and $\bar{\omega}_{\beta _{1}\gamma _{1}}=-\bar{\omega}_{\gamma _{1}\beta
_{1}}$ a solution to
\begin{equation}
Z_{\;\;\beta _{2}}^{\beta _{1}}\bar{\omega}_{\beta _{1}\gamma _{1}}\approx 0.
\label{a1}
\end{equation}%
We define an antisymmetric matrix $\hat{\omega}^{\alpha _{1}\beta _{1}}$
through the relation
\begin{equation}
\hat{\omega}^{\alpha _{1}\beta _{1}}\bar{\omega}_{\beta _{1}\gamma
_{1}}\approx \delta _{\;\;\gamma _{1}}^{\alpha _{1}}-Z_{\;\;\alpha
_{2}}^{\alpha _{1}}\bar{A}_{\gamma _{1}}^{\;\;\alpha _{2}}\equiv
D_{\;\;\gamma _{1}}^{\alpha _{1}}.  \label{a3}
\end{equation}%
Taking (\ref{a1}) into account, it results that $\hat{\omega}%
^{\alpha _{1}\beta _{1}}$ contains some ambiguities, namely it is
defined up to the transformation
\begin{equation}
\hat{\omega}^{\alpha _{1}\beta _{1}}\rightarrow \hat{\omega}^{\alpha
_{1}\beta _{1}}+Z_{\;\;\alpha _{2}}^{\alpha _{1}}q^{\alpha _{2}\beta
_{2}}Z_{\;\;\beta _{2}}^{\beta _{1}},  \label{az}
\end{equation}%
with $q^{\alpha _{2}\beta _{2}}$ some arbitrary, antisymmetric functions. On
the other hand, simple computation shows that the matrix $D_{\;\;\gamma
_{1}}^{\alpha _{1}}$ satisfies the properties
\begin{eqnarray}
\bar{A}_{\alpha _{1}}^{\;\;\alpha _{2}}D_{\;\;\gamma _{1}}^{\alpha _{1}}
&\approx &0,\;Z_{\;\;\gamma _{2}}^{\gamma _{1}}D_{\;\;\gamma _{1}}^{\alpha
_{1}}\approx 0,  \label{ax} \\
Z_{\;\;\alpha _{1}}^{\alpha _{0}}D_{\;\;\gamma _{1}}^{\alpha _{1}} &\approx
&Z_{\;\;\gamma _{1}}^{\alpha _{0}},\;D_{\;\;\gamma _{1}}^{\alpha
_{1}}D_{\;\;\lambda _{1}}^{\gamma _{1}}\approx D_{\;\;\lambda _{1}}^{\alpha
_{1}}.  \label{ay}
\end{eqnarray}%
Based on the latter formula from (\ref{ax}) we infer an alternative
expression for $D_{\;\;\gamma _{1}}^{\alpha _{1}}$, namely
\begin{equation}
D_{\;\;\gamma _{1}}^{\alpha _{1}}\approx \bar{A}_{\alpha _{0}}^{\;\;\alpha
_{1}}Z_{\;\;\gamma _{1}}^{\alpha _{0}},  \label{1qa}
\end{equation}%
for some functions $\bar{A}_{\alpha _{0}}^{\;\;\alpha _{1}}$. From the
former relation in (\ref{ay}) and (\ref{1qa}) we deduce that
\begin{equation}
Z_{\;\;\gamma _{1}}^{\gamma _{0}}D_{\;\;\gamma _{0}}^{\alpha _{0}}\approx 0,
\label{12b}
\end{equation}%
where
\begin{equation}
D_{\;\;\gamma _{0}}^{\alpha _{0}}\approx \delta _{\;\;\gamma _{0}}^{\alpha
_{0}}-Z_{\;\;\alpha _{1}}^{\alpha _{0}}\bar{A}_{\gamma _{0}}^{\;\;\alpha
_{1}}.  \label{11f}
\end{equation}%
At this stage, we can rewrite the Dirac bracket (\ref{11b}) without
separating a specific subset of independent constraints. In view of
this, we introduce an antisymmetric matrix $M^{\left( 2\right)
\alpha _{0}\beta _{0}}$ through the relation
\begin{equation}
M^{\left( 2\right) \alpha _{0}\beta _{0}}C_{\beta _{0}\gamma _{0}}^{\left(
2\right) }\approx D_{\;\;\gamma _{0}}^{\alpha _{0}},  \label{11c}
\end{equation}%
such that formula
\begin{equation}
\left[ F,G\right] ^{\left( 2\right) \ast }=\left[ F,G\right] -\left[ F,\chi
_{\alpha _{0}}\right] M^{\left( 2\right) \alpha _{0}\beta _{0}}\left[ \chi
_{\beta _{0}},G\right]  \label{14q}
\end{equation}%
defines the same Dirac bracket like (\ref{11b}) on the surface (\ref{1}). It
is simple to see that $M^{\left( 2\right) \alpha _{0}\beta _{0}}$ also
contains some ambiguities, being defined up to the transformation
\begin{equation}
M^{\left( 2\right) \alpha _{0}\beta _{0}}\rightarrow M^{\left( 2\right)
\alpha _{0}\beta _{0}}+Z_{\;\;\alpha _{1}}^{\alpha _{0}}\hat{q}^{\alpha
_{1}\beta _{1}}Z_{\;\;\beta _{1}}^{\beta _{0}},  \label{14r}
\end{equation}%
with $\hat{q}^{\alpha _{1}\beta _{1}}$ some antisymmetric, but otherwise
arbitrary functions. Relations (\ref{11x}) and (\ref{12b}) ensure that
\begin{equation}
\mathrm{rank}\left( D_{\;\;\gamma _{0}}^{\alpha _{0}}\right) \approx
M_{0}-M_{1}+M_{2},  \label{12a}
\end{equation}%
so the rank of $M^{\left( 2\right) \alpha _{0}\beta _{0}}C_{\beta _{0}\gamma
_{0}}^{\left( 2\right) }$ is equal to the number of independent second-class
constraints also in the presence of the second-order reducibility. At the
same time, we have that
\begin{equation}
\left[ \chi _{\alpha _{0}},G\right] ^{\left( 2\right) \ast }\approx -\bar{A}%
_{\alpha _{0}}^{\;\;\alpha _{1}}\left[ Z_{\;\;\alpha _{1}}^{\beta _{0}},G%
\right] \chi _{\beta _{0}},  \label{q12}
\end{equation}%
so we recover the property $\left[ \chi _{\alpha _{0}},G\right] ^{\left(
2\right) \ast }=0$ (for any $G$) on the surface of second-order reducible
second-class constraints. The fact the Dirac bracket given by (\ref{14q})
satisfies Jacobi's identity can be proved like in the first-order reducible
case. The analogous of the key relation (\ref{c5}) from the first-order
reducible situation is now $D_{\;\;\gamma _{0}}^{\alpha _{0}}\chi _{\alpha
_{0}}=\chi _{\gamma _{0}}$.

\subsubsection{Dirac bracket in terms of an invertible matrix}

Before expressing the Dirac bracket in terms of an invertible matrix, we
will analyze equations (\ref{a2}) and (\ref{a1}). The solution to (\ref{a2})
can be written as
\begin{equation}
\bar{A}_{\alpha _{1}}^{\;\;\alpha _{2}}\approx \bar{D}_{\;\;\lambda
_{2}}^{\alpha _{2}}A_{\alpha _{1}}^{\;\;\lambda _{2}},  \label{a4}
\end{equation}%
where $A_{\alpha _{1}}^{\;\;\lambda _{2}}$ are some functions chosen such
that the matrix
\begin{equation}
D_{\;\;\beta _{2}}^{\lambda _{2}}=Z_{\;\;\beta _{2}}^{\alpha _{1}}A_{\alpha
_{1}}^{\;\;\lambda _{2}}  \label{a5}
\end{equation}%
is of maximum rank,
\begin{equation}
\mathrm{rank}\left( D_{\;\;\beta _{2}}^{\lambda _{2}}\right) =M_{2},
\label{a6}
\end{equation}%
with $\bar{D}_{\;\;\lambda _{2}}^{\alpha _{2}}$ the inverse of $D_{\;\;\beta
_{2}}^{\lambda _{2}}$\footnote{%
Strictly speaking, the solution to (\ref{a2}) has the general form $\bar{A}%
_{\alpha _{1}}^{\;\;\alpha _{2}}\approx \bar{D}_{\;\;\lambda _{2}}^{\alpha
_{2}}A_{\alpha _{1}}^{\;\;\lambda _{2}}+u_{\;\;\alpha _{0}}^{\alpha
_{2}}Z_{\;\;\alpha _{1}}^{\alpha _{0}}+v^{\alpha _{2}\lambda _{1}}\bar{\omega%
}_{\lambda _{1}\alpha _{1}}$, where $u_{\;\;\alpha _{0}}^{\alpha _{2}}$ and $%
v^{\alpha _{2}\lambda _{1}}$ are arbitrary functions. By making the
redefinitions $u_{\;\;\alpha _{0}}^{\alpha
_{2}}=\bar{D}_{\;\;\lambda _{2}}^{\alpha _{2}}\hat{u}_{\;\;\alpha
_{0}}^{\lambda _{2}}$ and $\;v^{\alpha _{2}\lambda
_{1}}=\bar{D}_{\;\;\lambda _{2}}^{\alpha _{2}}\hat{v}^{\lambda
_{2}\lambda _{1}}$, with $\hat{u}_{\;\;\alpha _{0}}^{\lambda _{2}}$ and $%
\hat{v}^{\lambda _{2}\lambda _{1}}$ arbitrary, we can set $\bar{A}_{\alpha
_{1}}^{\;\;\alpha _{2}}$ in the form $\bar{A}_{\alpha _{1}}^{\;\;\alpha
_{2}}\approx \bar{D}_{\;\;\lambda _{2}}^{\alpha _{2}}\left( A_{\alpha
_{1}}^{\;\;\lambda _{2}}+\hat{u}_{\;\;\alpha _{0}}^{\lambda
_{2}}Z_{\;\;\alpha _{1}}^{\alpha _{0}}+\hat{v}^{\lambda _{2}\lambda _{1}}%
\bar{\omega}_{\lambda _{1}\alpha _{1}}\right) $. On the other hand, the
functions $A_{\alpha _{1}}^{\;\;\lambda _{2}}$ with the property that the
rank of matrix (\ref{a5}) is maximum are defined up to the transformation $%
A_{\alpha _{1}}^{\;\;\lambda _{2}}\rightarrow A_{\alpha _{1}}^{\prime
\;\;\lambda _{2}}=A_{\alpha _{1}}^{\;\;\lambda _{2}}+\tau _{\;\;\alpha
_{0}}^{\lambda _{2}}Z_{\;\;\alpha _{1}}^{\alpha _{0}}+\lambda ^{\lambda
_{2}\lambda _{1}}\bar{\omega}_{\lambda _{1}\alpha _{1}}$, in the sense that $%
Z_{\;\;\beta _{2}}^{\alpha _{1}}A_{\alpha _{1}}^{\;\;\lambda _{2}}\approx
Z_{\;\;\beta _{2}}^{\alpha _{1}}A_{\alpha _{1}}^{\prime \;\;\lambda _{2}}$,
where $\tau _{\;\;\alpha _{0}}^{\lambda _{2}}$ and $\lambda ^{\lambda
_{2}\lambda _{1}}$ are also arbitrary. Thus, we can always absorb the
quantity $\hat{u}_{\;\;\alpha _{0}}^{\lambda _{2}}Z_{\;\;\alpha
_{1}}^{\alpha _{0}}+\hat{v}^{\lambda _{2}\lambda _{1}}\bar{\omega}_{\lambda
_{1}\alpha _{1}}$ from $\bar{A}_{\alpha _{1}}^{\;\;\alpha _{2}}$ by
redefining $A_{\alpha _{1}}^{\;\;\lambda _{2}}$, such that we finally obtain
solution (\ref{a4}).}. Then, on the one hand we have that
\begin{equation}
D_{\;\;\gamma _{1}}^{\alpha _{1}}\approx \delta _{\;\;\gamma _{1}}^{\alpha
_{1}}-Z_{\;\;\alpha _{2}}^{\alpha _{1}}\bar{D}_{\;\;\lambda _{2}}^{\alpha
_{2}}A_{\gamma _{1}}^{\;\;\lambda _{2}}  \label{a7}
\end{equation}%
and on the other hand (inserting (\ref{a4}) in the former relation from (\ref%
{ax})) we can write
\begin{equation}
A_{\alpha _{1}}^{\;\;\sigma _{2}}D_{\;\;\gamma _{1}}^{\alpha _{1}}\approx 0.
\label{a8}
\end{equation}%
Substituting (\ref{1qa}) in (\ref{a8}), we are led to
\begin{equation}
\bar{A}_{\alpha _{0}}^{\;\;\alpha _{1}}A_{\alpha _{1}}^{\;\;\alpha
_{2}}\approx 0,  \label{12k}
\end{equation}%
which further implies
\begin{equation}
\bar{A}_{\alpha _{0}}^{\;\;\gamma _{1}}D_{\;\;\gamma _{1}}^{\alpha
_{1}}\approx \bar{A}_{\alpha _{0}}^{\;\;\alpha _{1}}.  \label{a9}
\end{equation}%
Based on the latter formula from (\ref{ax}), we find that the solution to (%
\ref{a1}) can be expressed as
\begin{equation}
\bar{\omega}_{\beta _{1}\gamma _{1}}\approx D_{\;\;\beta _{1}}^{\tau _{1}}%
\tilde{\omega}_{\tau _{1}\lambda _{1}}D_{\;\;\gamma _{1}}^{\lambda _{1}},
\label{a15}
\end{equation}%
where $\tilde{\omega}_{\tau _{1}\lambda _{1}}$ is antisymmetric. Acting with
$A_{\alpha _{1}}^{\;\;\alpha _{2}}$ on (\ref{a3}) and taking into account (%
\ref{a8}) and (\ref{a15}), we reach the equation
\begin{equation}
A_{\alpha _{1}}^{\;\;\alpha _{2}}\hat{\omega}^{\alpha _{1}\beta _{1}}\bar{%
\omega}_{\beta _{1}\gamma _{1}}\approx 0,  \label{a16}
\end{equation}%
whose solution can be chosen as\footnote{%
In fact, the general solution of (\ref{a16}) is given by $\hat{\omega}%
^{\alpha _{1}\beta _{1}}=D_{\;\;\rho _{1}}^{\alpha _{1}}\tilde{\omega}^{\rho
_{1}\sigma _{1}}D_{\;\;\sigma _{1}}^{\beta _{1}}+Z_{\;\;\alpha _{2}}^{\alpha
_{1}}u^{\alpha _{2}\beta _{2}}Z_{\;\;\beta _{2}}^{\beta _{1}}$, with $%
u^{\alpha _{2}\beta _{2}}$ arbitrary, antisymmetric functions. Since $\hat{%
\omega}^{\alpha _{1}\beta _{1}}$ are defined up to transformation (\ref{az}%
), we can always absorb the terms $Z_{\;\;\alpha _{2}}^{\alpha
_{1}}u^{\alpha _{2}\beta _{2}}Z_{\;\;\beta _{2}}^{\beta _{1}}$ through a
redefinition of $\hat{\omega}^{\alpha _{1}\beta _{1}}$ and finally arrive at
(\ref{a17}).}
\begin{equation}
\hat{\omega}^{\alpha _{1}\beta _{1}}=D_{\;\;\rho _{1}}^{\alpha _{1}}\tilde{%
\omega}^{\rho _{1}\sigma _{1}}D_{\;\;\sigma _{1}}^{\beta _{1}},  \label{a17}
\end{equation}%
with $\tilde{\omega}^{\rho _{1}\sigma _{1}}$ antisymmetric. With the
help of (\ref{a8}) and (\ref{a17}), it is easy to see that
\begin{equation}
A_{\alpha _{1}}^{\;\;\alpha _{2}}\hat{\omega}^{\alpha _{1}\beta _{1}}\approx
0.  \label{am}
\end{equation}%
Except from being antisymmetric, the matrices $\tilde{\omega}_{\tau
_{1}\lambda _{1}}$ and $\tilde{\omega}^{\rho _{1}\sigma _{1}}$ are arbitrary
at this point. Nevertheless, they can be chosen to satisfy a series of
useful properties, as the next theorem proves.

\begin{theorem}
\label{prop}The matrices of elements $\tilde{\omega}_{\tau _{1}\lambda _{1}}$
and $\tilde{\omega}^{\rho _{1}\sigma _{1}}$ can always be taken to satisfy
the following properties:

\noindent (a) (weak) invertibility,

\noindent (b) fulfillment of relation
\begin{equation}
\tilde{\omega}^{\rho _{1}\sigma _{1}}D_{\;\;\sigma _{1}}^{\beta _{1}}\tilde{%
\omega}_{\beta _{1}\lambda _{1}}\approx D_{\;\;\lambda _{1}}^{\rho
_{1}}, \label{a18}
\end{equation}

\noindent (c) (weak) mutual invertibility
\begin{equation}
\tilde{\omega}^{\rho _{1}\sigma _{1}}\tilde{\omega}_{\sigma _{1}\lambda
_{1}}\approx \delta _{\;\;\lambda _{1}}^{\rho _{1}}.  \label{a18a}
\end{equation}
\end{theorem}

\proof%
(a) Replacing the latter formula from (\ref{ay}) in (\ref{a15}) and (\ref%
{a17}), we infer the relations
\begin{eqnarray}
D_{\;\;\beta _{1}}^{\tau _{1}}\bar{\omega}_{\tau _{1}\lambda
_{1}}D_{\;\;\gamma _{1}}^{\lambda _{1}} &\approx &D_{\;\;\beta _{1}}^{\tau
_{1}}\tilde{\omega}_{\tau _{1}\lambda _{1}}D_{\;\;\gamma _{1}}^{\lambda
_{1}},  \label{a19} \\
D_{\;\;\rho _{1}}^{\alpha _{1}}\hat{\omega}^{\rho _{1}\sigma
_{1}}D_{\;\;\sigma _{1}}^{\beta _{1}} &\approx &D_{\;\;\rho _{1}}^{\alpha
_{1}}\tilde{\omega}^{\rho _{1}\sigma _{1}}D_{\;\;\sigma _{1}}^{\beta _{1}},
\label{a20}
\end{eqnarray}%
with the help of which we further deduce
\begin{eqnarray}
\tilde{\omega}_{\tau _{1}\lambda _{1}} &\approx &\bar{\omega}_{\tau
_{1}\lambda _{1}}+\bar{D}_{\;\;\tau _{2}}^{\sigma _{2}}A_{\tau
_{1}}^{\;\;\tau _{2}}\omega _{\sigma _{2}\gamma _{2}}A_{\lambda
_{1}}^{\;\;\lambda _{2}}\bar{D}_{\;\;\lambda _{2}}^{\gamma _{2}},
\label{a21} \\
\tilde{\omega}^{\rho _{1}\sigma _{1}} &\approx &\hat{\omega}^{\rho
_{1}\sigma _{1}}+Z_{\;\;\alpha _{2}}^{\rho _{1}}\omega ^{\alpha _{2}\beta
_{2}}Z_{\;\;\beta _{2}}^{\sigma _{1}},  \label{a22}
\end{eqnarray}%
for some antisymmetric matrices $\omega _{\sigma _{2}\gamma _{2}}$ and $%
\omega ^{\alpha _{2}\beta _{2}}$, taken to be invertible. Each of
the terms from the right-hand sides of formulae (\ref{a21}) and
(\ref{a22}) display null vectors. The null vectors of
$\bar{\omega}_{\tau _{1}\lambda _{1}}$ and $\hat{\omega}^{\rho
_{1}\sigma _{1}}$ are $Z_{\;\;\alpha _{2}}^{\lambda _{1}} $ and
$A_{\rho _{1}}^{\;\;\rho _{2}}$ respectively (see (\ref{a1}) and
(\ref{am}))\footnote{%
The most general form of the null vectors of the matrices $\bar{\omega}%
_{\tau _{1}\lambda _{1}}$ and $\hat{\omega}^{\rho _{1}\sigma _{1}}$ is $%
Z_{\;\;\alpha _{2}}^{\lambda _{1}}\nu ^{\alpha _{2}}$ and $%
A_{\rho _{1}}^{\;\;\rho _{2}}\xi _{\rho _{2}}$ respectively, with
$\nu ^{\alpha _{2}}$ and $\xi _{\rho _{2}}$ some arbitrary
functions, but this does not affect our proof.}, while the null
vectors of $\bar{D}_{\;\;\tau _{2}}^{\sigma _{2}}A_{\tau
_{1}}^{\;\;\tau _{2}}\omega _{\sigma _{2}\gamma _{2}}A_{\lambda
_{1}}^{\;\;\lambda _{2}}\bar{D}_{\;\;\lambda _{2}}^{\gamma _{2}}$ and $%
Z_{\;\;\alpha _{2}}^{\rho _{1}}\omega ^{\alpha _{2}\beta _{2}}Z_{\;\;\beta
_{2}}^{\sigma _{1}}$ are given by $\bar{A}_{\lambda _{0}}^{\;\;\lambda _{1}}$
and respectively $Z_{\;\;\sigma _{1}}^{\sigma _{0}}$. For this reason, the
only candidates for null vectors of $\tilde{\omega}_{\tau _{1}\lambda _{1}}$
and $\tilde{\omega}^{\rho _{1}\sigma _{1}}$ are on the one hand $%
Z_{\;\;\alpha _{2}}^{\lambda _{1}}$ and $A_{\rho _{1}}^{\;\;\rho
_{2}}$ respectively and on the other hand $\bar{A}_{\lambda
_{0}}^{\;\;\lambda _{1}}$ and $Z_{\;\;\sigma _{1}}^{\sigma _{0}} $
respectively. We show that none of these candidates are null
vectors. Indeed, from (\ref{a21}) and (\ref{a22}) we find

\begin{eqnarray}
Z_{\;\;\alpha _{2}}^{\lambda _{1}}\tilde{\omega}_{\tau _{1}\lambda _{1}}
&\approx &\bar{D}_{\;\;\tau _{2}}^{\sigma _{2}}A_{\tau _{1}}^{\;\;\tau
_{2}}\omega _{\sigma _{2}\alpha _{2}}\approx \bar{A}_{\tau _{1}}^{\;\;\sigma
_{2}}\omega _{\sigma _{2}\alpha _{2}},  \label{a23} \\
A_{\rho _{1}}^{\;\;\rho _{2}}\tilde{\omega}^{\rho _{1}\sigma _{1}} &\approx
&D_{\;\;\alpha _{2}}^{\rho _{2}}\omega ^{\alpha _{2}\beta _{2}}Z_{\;\;\beta
_{2}}^{\sigma _{1}}.  \label{a24}
\end{eqnarray}%
Since $D_{\;\;\alpha _{2}}^{\rho _{2}}$, $\omega _{\sigma _{2}\alpha
_{2}}$ and $\omega ^{\alpha _{2}\beta _{2}}$ are invertible, they
have no nontrivial null vectors. On the other hand, the matrix
$Z_{\;\;\beta _{2}}^{\sigma _{1}}\bar{A}_{\sigma _{1}}^{\;\;\sigma
_{2}}$ is of maximum rank (see (\ref{a2})), so neither
$\bar{A}_{\tau _{1}}^{\;\;\sigma _{2}}$ nor $Z_{\;\;\beta
_{2}}^{\sigma _{1}}$ can display nontrivial null vectors (i.e. there
are no nontrivial functions $\theta _{\sigma _{2}}$ or $\pi ^{\beta
_{2}}$ such that $\bar{A}_{\tau _{1}}^{\;\;\sigma _{2}}\theta
_{\sigma _{2}}\approx 0$ or $Z_{\;\;\beta _{2}}^{\sigma _{1}}\pi
^{\beta _{2}}\approx 0$). In consequence, the objects $Z_{\;\;\alpha
_{2}}^{\lambda _{1}}\tilde{\omega}_{\tau _{1}\lambda _{1}}$ and
$A_{\rho _{1}}^{\;\;\rho _{2}}\tilde{\omega}^{\rho _{1}\sigma _{1}}$
from (\ref{a23}) and (\ref{a24}) cannot vanish, and therefore the
matrices $\tilde{\omega}_{\tau _{1}\lambda _{1}}$ and
$\tilde{\omega}^{\rho _{1}\sigma _{1}}$ do not have the functions
$Z_{\;\;\alpha _{2}}^{\lambda _{1}}$ and $\bar{A}_{\rho
_{1}}^{\;\;\rho _{2}}$ as null vectors respectively. Multiplying (\ref{a21}) and (\ref%
{a22}) by $\bar{A}_{\lambda _{0}}^{\;\;\lambda _{1}}$ and $%
Z_{\;\;\sigma _{1}}^{\sigma _{0}}$ respectively, we infer the
relations
\begin{eqnarray}
\tilde{\omega}_{\tau _{1}\lambda _{1}}\bar{A}_{\lambda _{0}}^{\;\;\lambda
_{1}} &\approx &\bar{\omega}_{\tau _{1}\lambda _{1}}\bar{A}_{\lambda
_{0}}^{\;\;\lambda _{1}},  \label{b25} \\
\tilde{\omega}^{\rho _{1}\sigma _{1}}Z_{\;\;\sigma _{1}}^{\sigma _{0}}
&\approx &\hat{\omega}^{\rho _{1}\sigma _{1}}Z_{\;\;\sigma _{1}}^{\sigma
_{0}}.  \label{b26}
\end{eqnarray}%
The right-hand sides of (\ref{b25}) and (\ref{b26}) vanish for
\begin{eqnarray}
\bar{\omega}_{\tau _{1}\lambda _{1}} &=&A_{\tau _{1}}^{\;\;\sigma _{2}}\bar{%
\varepsilon}_{\sigma _{2}\gamma _{2}}A_{\lambda _{1}}^{\;\;\gamma _{2}},
\label{c250} \\
\hat{\omega}^{\rho _{1}\sigma _{1}} &=&Z_{\;\;\alpha _{2}}^{\rho _{1}}\hat{%
\varepsilon}^{\alpha _{2}\beta _{2}}Z_{\;\;\beta _{2}}^{\sigma _{1}},
\label{c260}
\end{eqnarray}%
where $\bar{\varepsilon}_{\sigma _{2}\gamma _{2}}$ and $\hat{\varepsilon}%
^{\alpha _{2}\beta _{2}}$ are antisymmetric. It is simple to see that $\bar{%
\omega}_{\tau _{1}\lambda _{1}}$ and $\hat{\omega}^{\rho _{1}\sigma
_{1}}$ given by (\ref{c250}) and (\ref{c260}) cannot be brought to
the form expressed by relations (\ref{a15}) and (\ref{a17})
respectively for any
choice of $\bar{\varepsilon}_{\sigma _{2}\gamma _{2}}$ or $\hat{\varepsilon}%
^{\alpha _{2}\beta _{2}}$. Thus, it follows that relations (\ref{c250}) and (%
\ref{c260}) cannot hold, such that $\bar{\omega}_{\tau _{1}\lambda _{1}}\bar{%
A}_{\lambda _{0}}^{\;\;\lambda _{1}}$ and $\hat{\omega}^{\rho _{1}\sigma
_{1}}Z_{\;\;\sigma _{1}}^{\sigma _{0}}$ do not vanish. Therefore, neither $%
\tilde{\omega}_{\tau _{1}\lambda _{1}}$ nor $\tilde{\omega}^{\rho
_{1}\sigma _{1}}$ (expressed by (\ref{a21}) and (\ref{a22})
respectively) have the
functions $\bar{A}_{\lambda _{0}}^{\;\;\lambda _{1}}$ and $%
Z_{\;\;\sigma _{1}}^{\sigma _{0}}$ as null vectors respectively, so
they are invertible. This proves (a).

(b) By straightforward computation, it results
\begin{eqnarray}
\tilde{\omega}^{\rho _{1}\sigma _{1}}D_{\;\;\sigma _{1}}^{\beta _{1}}
&\approx &\hat{\omega}^{\rho _{1}\beta _{1}},  \label{a27} \\
\hat{\omega}^{\rho _{1}\beta _{1}}\tilde{\omega}_{\beta _{1}\lambda _{1}}
&\approx &\hat{\omega}^{\rho _{1}\beta _{1}}\bar{\omega}_{\beta _{1}\lambda
_{1}}\approx D_{\;\;\lambda _{1}}^{\rho _{1}},  \label{a28}
\end{eqnarray}%
and hence
\begin{equation}
\tilde{\omega}^{\rho _{1}\sigma _{1}}D_{\;\;\sigma _{1}}^{\beta _{1}}\tilde{%
\omega}_{\beta _{1}\lambda _{1}}\approx D_{\;\;\lambda _{1}}^{\rho _{1}},
\label{a29}
\end{equation}%
which proves (b).

(c) Taking into account formulae (\ref{a1}), (\ref{a3}) and
(\ref{am}), from relations (\ref{a21}) and (\ref{a22}) we find
\begin{equation}
\tilde{\omega}^{\rho _{1}\sigma _{1}}\tilde{\omega}_{\sigma _{1}\lambda
_{1}}\approx D_{\;\;\lambda _{1}}^{\rho _{1}}+Z_{\;\;\alpha _{2}}^{\rho
_{1}}\omega ^{\alpha _{2}\beta _{2}}\omega _{\beta _{2}\gamma
_{2}}A_{\lambda _{1}}^{\;\;\lambda _{2}}\bar{D}_{\;\;\lambda _{2}}^{\gamma
_{2}}.  \label{a25}
\end{equation}%
Now, we take the matrices $\omega _{\sigma _{2}\gamma _{2}}$ and $\omega
^{\alpha _{2}\beta _{2}}$ to be mutually inverse, namely
\begin{equation}
\omega ^{\alpha _{2}\beta _{2}}\omega _{\beta _{2}\gamma _{2}}\approx \delta
_{\;\;\gamma _{2}}^{\alpha _{2}}.  \label{a26}
\end{equation}%
Substituting (\ref{a26}) into (\ref{a25}) and recalling formula
(\ref{a7}), we deduce (\ref{a18a}). This proves (c).$\Box $

With these elements at hand, the next theorem is shown to hold.

\begin{theorem}
\label{invert}There exists an invertible, antisymmetric matrix $\mu ^{\left(
2\right) \alpha _{0}\beta _{0}}$, in terms of which the Dirac bracket (\ref%
{14q}) becomes
\begin{equation}
\left[ F,G\right] ^{\left( 2\right) \ast }=\left[ F,G\right] -\left[ F,\chi
_{\alpha _{0}}\right] \mu ^{\left( 2\right) \alpha _{0}\beta _{0}}\left[
\chi _{\beta _{0}},G\right]  \label{24}
\end{equation}%
on the surface (\ref{1}).
\end{theorem}

\proof
First, we observe that $D_{\;\;\gamma _{0}}^{\alpha _{0}}$ given in (\ref%
{11f}) is a projector
\begin{equation}
D_{\;\;\gamma _{0}}^{\alpha _{0}}D_{\;\;\lambda _{0}}^{\gamma _{0}}\approx
D_{\;\;\lambda _{0}}^{\alpha _{0}}  \label{15}
\end{equation}%
and satisfies the relations
\begin{equation}
\bar{A}_{\alpha _{0}}^{\;\;\gamma _{1}}D_{\;\;\gamma _{0}}^{\alpha
_{0}}\approx 0,\;D_{\;\;\gamma _{0}}^{\alpha _{0}}\chi _{\alpha _{0}}\approx
\chi _{\gamma _{0}}.  \label{17}
\end{equation}%
Multiplying (\ref{11c}) by $\bar{A}_{\alpha _{0}}^{\;\;\gamma _{1}}$
and using (\ref{17}), we obtain the equation
\begin{equation}
\bar{A}_{\alpha _{0}}^{\;\;\gamma _{1}}M^{\left( 2\right) \alpha _{0}\beta
_{0}}C_{\beta _{0}\gamma _{0}}^{\left( 2\right) }\approx 0,  \label{pp1}
\end{equation}%
which then leads to
\begin{equation}
\bar{A}_{\alpha _{0}}^{\;\;\gamma _{1}}M^{\left( 2\right) \alpha _{0}\beta
_{0}}\approx f^{\gamma _{1}\beta _{1}}Z_{\;\;\beta _{1}}^{\beta _{0}},
\label{19}
\end{equation}%
for some functions $f^{\gamma _{1}\beta _{1}}$. Acting with
$D_{\;\;\beta _{0}}^{\tau _{0}}$ on (\ref{19}) and taking into
account (\ref{12b}), we reach the relation
\begin{equation}
\bar{A}_{\alpha _{0}}^{\;\;\gamma _{1}}M^{\left( 2\right) \alpha _{0}\beta
_{0}}D_{\;\;\beta _{0}}^{\tau _{0}}\approx 0,  \label{19x}
\end{equation}%
which combined with the former formula in (\ref{17}) produces
\begin{equation}
M^{\left( 2\right) \alpha _{0}\beta _{0}}D_{\;\;\beta _{0}}^{\tau
_{0}}\approx \lambda ^{\tau _{0}\beta _{0}}D_{\;\;\beta _{0}}^{\alpha _{0}},
\label{19y}
\end{equation}%
for some $\lambda ^{\tau _{0}\beta _{0}}$. Applying now $D_{\;\;\alpha
_{0}}^{\tau _{0}}$ on (\ref{11c}) and employing relation (\ref{19y}), we
deduce
\begin{equation}
-\lambda ^{\tau _{0}\alpha _{0}}D_{\;\;\alpha _{0}}^{\beta _{0}}C_{\beta
_{0}\gamma _{0}}^{\left( 2\right) }\approx D_{\;\;\gamma _{0}}^{\tau _{0}}.
\label{19w}
\end{equation}%
On the other hand, the latter formula from (\ref{17}) ensures that
\begin{equation}
D_{\;\;\alpha _{0}}^{\beta _{0}}C_{\beta _{0}\gamma _{0}}^{\left( 2\right)
}\approx C_{\alpha _{0}\gamma _{0}}^{\left( 2\right) },  \label{19q}
\end{equation}%
such that with the aid of the results expressed by (\ref{19w}) and
(\ref{19q}) we find
\begin{equation}
-\lambda ^{\tau _{0}\alpha _{0}}C_{\alpha _{0}\gamma _{0}}^{\left( 2\right)
}\approx D_{\;\;\gamma _{0}}^{\tau _{0}}.  \label{19z}
\end{equation}%
Comparing (\ref{19z}) with (\ref{11c}) and recalling that the elements $%
M^{\left( 2\right) \alpha _{0}\beta _{0}}$ are defined up to transformation (%
\ref{14r}), we infer the relation
\begin{equation}
M^{\left( 2\right) \tau _{0}\alpha _{0}}=-\lambda ^{\tau _{0}\alpha _{0}},
\label{19u}
\end{equation}%
which inserted in (\ref{19y}) provides the equation
\begin{equation}
D_{\;\;\alpha _{0}}^{\tau _{0}}M^{\left( 2\right) \alpha _{0}\beta
_{0}}\approx M^{\left( 2\right) \tau _{0}\alpha _{0}}D_{\;\;\alpha
_{0}}^{\beta _{0}}.  \label{19t}
\end{equation}%
Using once more the fact that the elements $M^{\left( 2\right) \alpha
_{0}\beta _{0}}$ are defined up to (\ref{14r}), from (\ref{19t}) it results
\begin{equation}
M^{\left( 2\right) \alpha _{0}\beta _{0}}\approx D_{\;\;\lambda
_{0}}^{\alpha _{0}}\mu ^{\left( 2\right) \lambda _{0}\sigma
_{0}}D_{\;\;\sigma _{0}}^{\beta _{0}},  \label{20}
\end{equation}%
where the elements $\mu ^{\left( 2\right) \lambda _{0}\sigma _{0}}$
define an antisymmetric matrix. Based on the former formula from
(\ref{17}) and on relation (\ref{20}), we infer
\begin{equation}
\bar{A}_{\alpha _{0}}^{\;\;\gamma _{1}}M^{\left( 2\right) \alpha _{0}\beta
_{0}}\approx 0.  \label{20y}
\end{equation}%
Replacing (\ref{15}) in (\ref{20}), we arrive at
\begin{equation}
D_{\;\;\lambda _{0}}^{\alpha _{0}}M^{\left( 2\right) \lambda _{0}\sigma
_{0}}D_{\;\;\sigma _{0}}^{\beta _{0}}\approx D_{\;\;\lambda _{0}}^{\alpha
_{0}}\mu ^{\left( 2\right) \lambda _{0}\sigma _{0}}D_{\;\;\sigma
_{0}}^{\beta _{0}},  \label{20x}
\end{equation}%
which leads to
\begin{equation}
\mu ^{\left( 2\right) \lambda _{0}\sigma _{0}}\approx M^{\left( 2\right)
\lambda _{0}\sigma _{0}}+Z_{\;\;\lambda _{1}}^{\lambda _{0}}\Omega ^{\lambda
_{1}\sigma _{1}}Z_{\;\;\sigma _{1}}^{\sigma _{0}},  \label{21}
\end{equation}%
for some antisymmetric functions $\Omega ^{\lambda _{1}\sigma _{1}}$. At
this point we show that the matrix $\mu ^{\left( 2\right) \lambda _{0}\sigma
_{0}}$ can indeed be taken to be invertible. If we choose $\Omega ^{\lambda
_{1}\sigma _{1}}$ as $\Omega ^{\lambda _{1}\sigma _{1}}=\tilde{\omega}%
^{\lambda _{1}\sigma _{1}}$, where $\tilde{\omega}^{\lambda _{1}\sigma _{1}}$
is precisely the invertible matrix given in (\ref{a22}), we get
\begin{equation}
\mu ^{\left( 2\right) \lambda _{0}\sigma _{0}}\approx M^{\left( 2\right)
\lambda _{0}\sigma _{0}}+Z_{\;\;\lambda _{1}}^{\lambda _{0}}\tilde{\omega}%
^{\lambda _{1}\sigma _{1}}Z_{\;\;\sigma _{1}}^{\sigma _{0}}.  \label{21z}
\end{equation}%
In the following, we show that the matrix of elements
\begin{equation}
\mu _{\sigma _{0}\rho _{0}}^{\left( 2\right) }\approx C_{\sigma _{0}\rho
_{0}}^{\left( 2\right) }+\bar{A}_{\sigma _{0}}^{\;\;\rho _{1}}\tilde{\omega}%
_{\rho _{1}\tau _{1}}\bar{A}_{\rho _{0}}^{\;\;\tau _{1}},  \label{21x}
\end{equation}%
with $\tilde{\omega}_{\rho _{1}\tau _{1}}$ the invertible matrix from (\ref%
{a21}), is nothing but the inverse of $\mu ^{\left( 2\right) \lambda
_{0}\sigma _{0}}$ expressed in (\ref{21z}). Indeed, relying on relations (%
\ref{11e}), (\ref{1qa}), (\ref{11c}) and (\ref{20y}), by direct
computation we find
\begin{equation}
\mu ^{\left( 2\right) \lambda _{0}\sigma _{0}}\mu _{\sigma _{0}\rho
_{0}}^{\left( 2\right) }\approx D_{\;\;\rho _{0}}^{\lambda
_{0}}+Z_{\;\;\lambda _{1}}^{\lambda _{0}}\tilde{\omega}^{\lambda _{1}\sigma
_{1}}D_{\;\;\sigma _{1}}^{\rho _{1}}\tilde{\omega}_{\rho _{1}\tau _{1}}\bar{A%
}_{\rho _{0}}^{\;\;\tau _{1}}.  \label{21y}
\end{equation}%
Employing Theorem \ref{prop} (see (\ref{a18})) and the former equation in (%
\ref{ay}), we deduce the relation
\begin{equation}
Z_{\;\;\lambda _{1}}^{\lambda _{0}}\tilde{\omega}^{\lambda _{1}\sigma
_{1}}D_{\;\;\sigma _{1}}^{\rho _{1}}\tilde{\omega}_{\rho _{1}\tau _{1}}\bar{A%
}_{\rho _{0}}^{\;\;\tau _{1}}\approx Z_{\;\;\lambda _{1}}^{\lambda
_{0}}D_{\;\;\tau _{1}}^{\lambda _{1}}\bar{A}_{\rho _{0}}^{\;\;\tau
_{1}}\approx Z_{\;\;\lambda _{1}}^{\lambda _{0}}\bar{A}_{\rho
_{0}}^{\;\;\lambda _{1}},  \label{21w}
\end{equation}%
which replaced in (\ref{21y}) reduces to
\begin{equation}
\mu ^{\left( 2\right) \lambda _{0}\sigma _{0}}\mu _{\sigma _{0}\rho
_{0}}^{\left( 2\right) }\approx \delta _{\;\;\rho _{0}}^{\lambda _{0}}.
\label{21q}
\end{equation}%
The above formula proves that the matrix of elements $\mu ^{\left( 2\right)
\lambda _{0}\sigma _{0}}$ from (\ref{21z}) is (weakly) invertible and
therefore completes the proof of this theorem. $\Box $

Formula (\ref{24}) plays a key role in what follows. It allows one to
express the original Dirac bracket (\ref{11b}), initially written only in
terms of a \emph{subset of independent} second-class constraint functions,
with the help of an invertible matrix, whose indices cover the \emph{whole
set} of reducible second-class constraints. Inspired by this result, we will
be able to find an irreducible second-class constraint set, whose Dirac
bracket is (weakly) equal to (\ref{24}).

\subsection{Irreducible approach}

\subsubsection{Intermediate system}

Now, we introduce some new variables, $\left( y_{\alpha _{1}}\right)
_{\alpha _{1}=1,\ldots ,M_{1}}$, independent of the original phase-space
variables $z^{a}$, with the Poisson brackets
\begin{equation}
\left[ y_{\alpha _{1}},y_{\beta _{1}}\right] =\omega _{\alpha _{1}\beta
_{1}},  \label{25}
\end{equation}%
where the elements $\omega _{\alpha _{1}\beta _{1}}$ define an invertible,
antisymmetric (but otherwise arbitrary) matrix\footnote{%
The elements $\omega _{\alpha _{1}\beta _{1}}$ may depend at most on the
newly added variables, just like the objects $\Gamma _{\alpha _{1}\beta
_{1}} $ from Section \ref{ford}.}, and consider the system subject to the
reducible second-class constraints
\begin{equation}
\chi _{\alpha _{0}}\approx 0,\;y_{\alpha _{1}}\approx 0.  \label{26}
\end{equation}%
The system subject to the second-class constraints (\ref{26}) will
be called an \emph{intermediate system} in what follows. The Dirac
bracket on the larger phase space, locally described by $\left(
z^{a},y_{\alpha _{1}}\right) $, corresponding to the above
second-class constraints reads as
\begin{equation}
\left. \left[ F,G\right] ^{\left( 2\right) \ast }\right\vert _{z,y}=\left[
F,G\right] -\left[ F,\chi _{\alpha _{0}}\right] \mu ^{\left( 2\right) \alpha
_{0}\beta _{0}}\left[ \chi _{\beta _{0}},G\right] -\left[ F,y_{\alpha _{1}}%
\right] \omega ^{\alpha _{1}\beta _{1}}\left[ y_{\beta _{1}},G\right] ,
\label{27}
\end{equation}%
where the Poisson brackets from the right-hand side of (\ref{27}) contain
derivatives with respect to all $z^{a}$'s and $y_{\alpha _{1}}$'s, and $%
\omega ^{\alpha _{1}\beta _{1}}$ denotes the elements of the inverse of $%
\omega _{\alpha _{1}\beta _{1}}$. On the one hand, the most general form of
a smooth function defined on the phase space with the local coordinates $%
\left( z^{a},y_{\alpha _{1}}\right) $ is
\begin{equation}
F\left( z^{a},y_{\alpha _{1}}\right) =F_{0}\left( z^{a}\right) +b^{\lambda
_{1}}\left( z^{a}\right) y_{\lambda _{1}}+b^{\lambda _{1}\rho _{1}}\left(
z^{a}\right) y_{\lambda _{1}}y_{\rho _{1}}+\cdots ,  \label{27tr}
\end{equation}%
for some smooth functions $b^{\lambda _{1}}\left( z^{a}\right) $, $%
b^{\lambda _{1}\rho _{1}}\left( z^{a}\right) $, etc. On the other hand,
direct computation yields
\begin{equation}
\left[ F,G\right] ^{\left( 2\right) \ast }\approx \left[ F_{0},G_{0}\right]
^{\left( 2\right) \ast },  \label{27qf}
\end{equation}%
where the previous weak equality is defined on the surface (\ref{26}).
Moreover, equations (\ref{1}) and (\ref{26}) describe the same surface, but
embedded in phase spaces of different dimensions. In other words, equations (%
\ref{1}) and (\ref{26}) are equivalent descriptions of the
\emph{same surface of constraints}. For this reason, we will employ
the same symbol of weak
equality for both descriptions\footnote{%
It is understood that if we work with functions defined on the phase
space of coordinates $z^{a}$, then we employ representation
(\ref{1}), but if we
work with functions of $\left( z^{a},y_{\alpha _{1}}\right) $, then we use (%
\ref{26}).}. Inserting (\ref{27tr}) in (\ref{27}) and taking
(\ref{27qf}) into account, we obtain
\begin{equation}
\left. \left[ F,G\right] ^{\left( 2\right) \ast }\right\vert _{z,y}\approx %
\left[ F,G\right] ^{\left( 2\right) \ast }.  \label{28}
\end{equation}%
We recall that the Dirac bracket $\left[ F,G\right] ^{\left( 2\right) \ast }$
contains only derivatives with respect to the original variables $z^{a}$.

Formula (\ref{28}) is important since together with (\ref{24}) it
opens the perspective towards the construction of an irreducible
second-class constraint system associated with the original,
second-order reducible one, but on the larger phase space $\left(
z^{a},y_{\alpha _{1}}\right) $.

\subsubsection{Irreducible system\label{irredsy}}

Now, we choose $\omega _{\gamma _{1}\lambda _{1}}$ from (\ref{25}) such that
\begin{equation}
\tilde{\omega}_{\alpha _{1}\beta _{1}}=\hat{E}_{\;\;\alpha _{1}}^{\gamma
_{1}}\omega _{\gamma _{1}\lambda _{1}}\hat{E}_{\;\;\beta _{1}}^{\lambda
_{1}},  \label{27q}
\end{equation}%
for an invertible matrix, of elements $\hat{E}_{\;\;\alpha _{1}}^{\gamma
_{1}}$, with the help of which we introduce the functions
\begin{equation}
A_{\sigma _{0}}^{\;\;\rho _{1}}=\hat{E}_{\;\;\alpha _{1}}^{\rho _{1}}\bar{A}%
_{\sigma _{0}}^{\;\;\alpha _{1}}.  \label{27y}
\end{equation}%
Then, we have that
\begin{equation}
\tilde{\omega}^{\alpha _{1}\beta _{1}}=\hat{e}_{\;\;\sigma _{1}}^{\alpha
_{1}}\omega ^{\sigma _{1}\tau _{1}}\hat{e}_{\;\;\tau _{1}}^{\beta _{1}},
\label{27w}
\end{equation}%
where $\hat{e}_{\;\;\sigma _{1}}^{\alpha _{1}}$ is the inverse of $\hat{E}%
_{\;\;\alpha _{1}}^{\gamma _{1}}$. By means of (\ref{27y}) we find
\begin{equation}
\bar{A}_{\sigma _{0}}^{\;\;\alpha _{1}}=A_{\sigma _{0}}^{\;\;\rho _{1}}\hat{e%
}_{\;\;\rho _{1}}^{\alpha _{1}}.  \label{qw21}
\end{equation}

In this context the following theorem can be shown to hold.

\begin{theorem}
\label{eleme}The elements $\hat{e}_{\;\;\sigma _{1}}^{\alpha _{1}}$ and $%
\hat{E}_{\;\;\beta _{1}}^{\tau _{1}}$ can always be taken such that
\begin{equation}
\hat{E}_{\;\;\sigma _{1}}^{\alpha _{1}}D_{\;\;\tau _{1}}^{\sigma _{1}}\hat{e}%
_{\;\;\beta _{1}}^{\tau _{1}}\approx D_{\;\;\beta _{1}}^{\alpha _{1}}.
\label{27qq}
\end{equation}
\end{theorem}

\proof
We choose $\hat{E}_{\;\;\beta _{1}}^{\alpha _{1}}$ such that
\begin{equation}
A_{\alpha _{0}}^{\;\;\alpha _{1}}=\sigma _{\alpha _{0}\beta _{0}}\sigma
^{\alpha _{1}\beta _{1}}Z_{\;\;\beta _{1}}^{\beta _{0}},  \label{27qw}
\end{equation}%
where $\sigma _{\alpha _{0}\beta _{0}}$ is invertible and $\sigma ^{\alpha
_{1}\beta _{1}}$ is invertible and symmetric. If we take
\begin{equation}
A_{\alpha _{1}}^{\;\;\alpha _{2}}=\sigma _{\alpha _{1}\lambda _{1}}\sigma
^{\alpha _{2}\beta _{2}}Z_{\;\;\beta _{2}}^{\lambda _{1}},  \label{27ww}
\end{equation}%
with $\sigma ^{\alpha _{2}\beta _{2}}$ invertible and $\sigma _{\alpha
_{1}\lambda _{1}}$ the inverse of $\sigma ^{\alpha _{1}\beta _{1}}$, then we
obtain that (\ref{a6}) is satisfied\footnote{%
With this choice of $A_{\alpha _{1}}^{\;\;\alpha _{2}}$, we have that $%
D_{\;\;\lambda _{2}}^{\alpha _{2}}=Z_{\;\;\lambda _{2}}^{\alpha _{1}}\sigma
_{\alpha _{1}\lambda _{1}}Z_{\;\;\beta _{2}}^{\lambda _{1}}\sigma ^{\alpha
_{2}\beta _{2}}$. Because $Z_{\;\;\lambda _{2}}^{\alpha _{1}}$ has no
nontrivial null vectors, it follows that the matrix of elements $%
Z_{\;\;\lambda _{2}}^{\alpha _{1}}\sigma _{\alpha _{1}\lambda
_{1}}Z_{\;\;\beta _{2}}^{\lambda _{1}}$ is invertible. On the other hand, $%
\sigma ^{\alpha _{2}\beta _{2}}$ is by hypothesis invertible, so $%
D_{\;\;\lambda _{2}}^{\alpha _{2}}$ is the same, as required by (\ref{a6}).}%
. Employing (\ref{27qw})--(\ref{27ww}) and recalling (\ref{11x}) we get
\begin{equation}
A_{\alpha _{1}}^{\;\;\alpha _{2}}A_{\alpha _{0}}^{\;\;\alpha _{1}}\approx 0.
\label{27qz}
\end{equation}%
Expressing the first-order reducibility functions from (\ref{27qw})--(\ref%
{27ww})
\begin{equation}
Z_{\;\;\alpha _{1}}^{\alpha _{0}}=\sigma ^{\alpha _{0}\beta _{0}}\sigma
_{\alpha _{1}\beta _{1}}A_{\beta _{0}}^{\;\;\beta _{1}},\;Z_{\;\;\lambda
_{2}}^{\lambda _{1}}=\sigma ^{\lambda _{1}\tau _{1}}\sigma _{\lambda
_{2}\tau _{2}}A_{\tau _{1}}^{\;\;\tau _{2}},  \label{27wz}
\end{equation}%
where $\sigma ^{\alpha _{0}\beta _{0}}$ and $\sigma _{\lambda _{2}\tau _{2}}$
are the inverses of $\sigma _{\alpha _{0}\beta _{0}}$ and respectively $%
\sigma ^{\alpha _{2}\beta _{2}}$, we deduce
\begin{equation}
Z_{\;\;\alpha _{1}}^{\alpha _{0}}\hat{e}_{\;\;\lambda _{1}}^{\alpha
_{1}}Z_{\;\;\lambda _{2}}^{\lambda _{1}}=\sigma ^{\alpha _{0}\beta
_{0}}\sigma _{\lambda _{2}\tau _{2}}A_{\beta _{0}}^{\;\;\beta _{1}}\sigma
_{\alpha _{1}\beta _{1}}\hat{e}_{\;\;\lambda _{1}}^{\alpha _{1}}\sigma
^{\lambda _{1}\tau _{1}}A_{\tau _{1}}^{\;\;\tau _{2}}.  \label{27qy}
\end{equation}%
Formula (\ref{27w}) can be rewritten as $\tilde{\omega}^{\alpha _{1}\beta
_{1}}=\hat{e}^{\alpha _{1}\sigma _{1}}\check{\omega}_{\sigma _{1}\tau _{1}}%
\hat{e}^{\beta _{1}\tau _{1}}$, with $\check{\omega}_{\sigma _{1}\tau
_{1}}=\sigma _{\sigma _{1}\rho _{1}}\omega ^{\rho _{1}\gamma _{1}}\sigma
_{\gamma _{1}\tau _{1}}$ and $\hat{e}^{\alpha _{1}\sigma _{1}}=\hat{e}%
_{\;\;\lambda _{1}}^{\alpha _{1}}\sigma ^{\lambda _{1}\sigma _{1}}$. Because
the matrix $\sigma _{\sigma _{1}\rho _{1}}$ is symmetric and $\omega ^{\rho
_{1}\gamma _{1}}$ antisymmetric, it follows that $\check{\omega}_{\sigma
_{1}\tau _{1}}$ is antisymmetric. The antisymmetry property of both $\tilde{%
\omega}^{\alpha _{1}\beta _{1}}$ and $\check{\omega}_{\sigma _{1}\tau _{1}}$
implies that the quantities $\hat{e}^{\alpha _{1}\sigma _{1}}$ can be taken
to be symmetric\footnote{%
The other possibility, namely the antisymmetry of $\hat{e}^{\alpha
_{1}\sigma _{1}}$, will not be considered in the sequel.}
\begin{equation}
\hat{e}^{\alpha _{1}\sigma _{1}}=\hat{e}_{\;\;\lambda _{1}}^{\alpha
_{1}}\sigma ^{\lambda _{1}\sigma _{1}}=\hat{e}^{\sigma _{1}\alpha _{1}}.
\label{27wy}
\end{equation}%
By means of (\ref{27wy}) we infer $\sigma _{\alpha _{1}\beta _{1}}\hat{e}%
_{\;\;\lambda _{1}}^{\alpha _{1}}\sigma ^{\lambda _{1}\tau _{1}}=\hat{e}%
_{\;\;\beta _{1}}^{\tau _{1}}$, such that from (\ref{27qy}) (and also (\ref%
{qw21})) we find the relation
\begin{equation}
Z_{\;\;\alpha _{1}}^{\alpha _{0}}\hat{e}_{\;\;\lambda _{1}}^{\alpha
_{1}}Z_{\;\;\lambda _{2}}^{\lambda _{1}}=\sigma ^{\alpha _{0}\beta
_{0}}\sigma _{\lambda _{2}\tau _{2}}\bar{A}_{\beta _{0}}^{\;\;\beta
_{1}}A_{\beta _{1}}^{\;\;\tau _{2}}.  \label{27qr}
\end{equation}%
Substituting now (\ref{12k}) in (\ref{27qr}) we obtain
\begin{equation}
Z_{\;\;\alpha _{1}}^{\alpha _{0}}\hat{e}_{\;\;\lambda _{1}}^{\alpha
_{1}}Z_{\;\;\lambda _{2}}^{\lambda _{1}}\approx 0.  \label{27qs}
\end{equation}%
With relations (\ref{27qz}) and (\ref{27qs}) at hand, we are in the position
to prove (\ref{27qq}). If we make the notation
\begin{equation}
\hat{D}_{\;\;\beta _{1}}^{\alpha _{1}}=\hat{e}_{\;\;\sigma _{1}}^{\alpha
_{1}}D_{\;\;\tau _{1}}^{\sigma _{1}}\hat{E}_{\;\;\beta _{1}}^{\tau _{1}},
\label{27wd}
\end{equation}%
then it is easy to see that $\hat{D}_{\;\;\beta _{1}}^{\alpha _{1}}$ is a
projector
\begin{equation}
\hat{D}_{\;\;\beta _{1}}^{\alpha _{1}}\hat{D}_{\;\;\lambda _{1}}^{\beta
_{1}}\approx \hat{D}_{\;\;\lambda _{1}}^{\alpha _{1}}.  \label{27xr}
\end{equation}%
On the other hand, with the aid of (\ref{27y}) and (\ref{27qz}) we deduce
\begin{equation}
\bar{A}_{\alpha _{0}}^{\;\;\beta _{1}}\hat{D}_{\;\;\beta _{1}}^{\alpha
_{1}}\approx \bar{A}_{\alpha _{0}}^{\;\;\alpha _{1}}.  \label{27ws}
\end{equation}%
Applying $Z_{\;\;\alpha _{1}}^{\alpha _{0}}$ on (\ref{27wd}) and using (\ref%
{27qs}) it follows
\begin{equation}
Z_{\;\;\alpha _{1}}^{\alpha _{0}}\hat{D}_{\;\;\beta _{1}}^{\alpha
_{1}}\approx Z_{\;\;\beta _{1}}^{\alpha _{0}}.  \label{27za}
\end{equation}%
Multiplying (\ref{27ws}) with $Z_{\;\;\rho _{1}}^{\alpha _{0}}$ and
respectively (\ref{27za}) with $\bar{A}_{\alpha _{0}}^{\;\;\alpha _{1}}$ we
reach the equations
\begin{equation}
\hat{D}_{\;\;\beta _{1}}^{\alpha _{1}}D_{\;\;\rho _{1}}^{\beta _{1}}\approx
D_{\;\;\rho _{1}}^{\alpha _{1}},\;D_{\;\;\beta _{1}}^{\alpha _{1}}\hat{D}%
_{\;\;\rho _{1}}^{\beta _{1}}\approx D_{\;\;\rho _{1}}^{\alpha _{1}}.
\label{27zb}
\end{equation}%
The general solution to equations (\ref{27zb}) can be represented like
\begin{equation}
\hat{D}_{\;\;\beta _{1}}^{\alpha _{1}}\approx D_{\;\;\beta _{1}}^{\alpha
_{1}}+Z_{\;\;\lambda _{2}}^{\alpha _{1}}M_{\;\;\tau _{2}}^{\lambda
_{2}}A_{\beta _{1}}^{\;\;\tau _{2}},  \label{27sr}
\end{equation}%
for some matrix $M_{\;\;\tau _{2}}^{\lambda _{2}}$. Direct computation shows
that
\begin{equation}
\hat{D}_{\;\;\beta _{1}}^{\alpha _{1}}\hat{D}_{\;\;\lambda _{1}}^{\beta
_{1}}\approx D_{\;\;\lambda _{1}}^{\alpha _{1}}+Z_{\;\;\lambda _{2}}^{\alpha
_{1}}M_{\;\;\tau _{2}}^{\lambda _{2}}D_{\;\;\beta _{2}}^{\tau
_{2}}M_{\;\;\rho _{2}}^{\beta _{2}}A_{\lambda _{1}}^{\;\;\rho _{2}}.
\label{27sy}
\end{equation}%
Comparing (\ref{27sy}) with (\ref{27xr}) and employing (\ref{27sr}) we find
that $M_{\;\;\tau _{2}}^{\lambda _{2}}$ are solutions to the equations
\begin{equation}
Z_{\;\;\lambda _{2}}^{\alpha _{1}}M_{\;\;\tau _{2}}^{\lambda
_{2}}D_{\;\;\beta _{2}}^{\tau _{2}}M_{\;\;\rho _{2}}^{\beta _{2}}A_{\lambda
_{1}}^{\;\;\rho _{2}}\approx Z_{\;\;\lambda _{2}}^{\alpha _{1}}M_{\;\;\tau
_{2}}^{\lambda _{2}}A_{\lambda _{1}}^{\;\;\tau _{2}}.  \label{27sg}
\end{equation}%
It is simple to see that equations (\ref{27sg}) possess two kinds of
solutions, namely
\begin{equation}
M_{\;\;\tau _{2}}^{\lambda _{2}}=0,  \label{27sm}
\end{equation}%
and respectively
\begin{equation}
M_{\;\;\tau _{2}}^{\lambda _{2}}=\bar{D}_{\;\;\tau _{2}}^{\lambda _{2}}.
\label{27sn}
\end{equation}%
If we take the second solution, (\ref{27sm})\footnote{%
Solution (\ref{27sn}) leads to the equation $\hat{e}_{\;\;\sigma
_{1}}^{\alpha _{1}}D_{\;\;\tau _{1}}^{\sigma _{1}}\hat{E}_{\;\;\beta
_{1}}^{\tau _{1}}\approx \delta _{\;\;\beta _{1}}^{\alpha _{1}}$. This
further provides the relation $D_{\;\;\beta _{1}}^{\sigma _{1}}\approx
\delta _{\;\;\beta _{1}}^{\alpha _{1}}$, which contradicts (\ref{a7}).},
from (\ref{27sr}) we obtain
\begin{equation}
\hat{D}_{\;\;\beta _{1}}^{\alpha _{1}}\approx D_{\;\;\beta _{1}}^{\alpha
_{1}},  \label{27pq}
\end{equation}%
which ensures (\ref{27qq}). This proves the theorem. $\Box $

Inserting (\ref{27q})--(\ref{27w}) in (\ref{a18}) and recalling (\ref{27qq})
it is easy to deduce the relation
\begin{equation}
\omega ^{\alpha _{1}\tau _{1}}D_{\;\;\tau _{1}}^{\sigma _{1}}\omega _{\sigma
_{1}\beta _{1}}\approx D_{\;\;\beta _{1}}^{\alpha _{1}}.  \label{27wp}
\end{equation}%
On the other hand, formulas (\ref{27q})--(\ref{27w}) indicate that $\mu
^{\left( 2\right) \lambda _{0}\sigma _{0}}$ and $\mu _{\sigma _{0}\rho
_{0}}^{\left( 2\right) }$ provided by (\ref{21z})--(\ref{21x}) take the form
\begin{eqnarray}
\mu ^{\left( 2\right) \lambda _{0}\sigma _{0}} &\approx &M^{\left( 2\right)
\lambda _{0}\sigma _{0}}+Z_{\;\;\lambda _{1}}^{\lambda _{0}}\hat{e}%
_{\;\;\sigma _{1}}^{\lambda _{1}}\omega ^{\sigma _{1}\tau _{1}}\hat{e}%
_{\;\;\tau _{1}}^{\gamma _{1}}Z_{\;\;\gamma _{1}}^{\sigma _{0}},  \label{27z}
\\
\mu _{\sigma _{0}\rho _{0}}^{\left( 2\right) } &\approx &C_{\sigma _{0}\rho
_{0}}^{\left( 2\right) }+A_{\sigma _{0}}^{\;\;\rho _{1}}\omega _{\rho
_{1}\tau _{1}}A_{\rho _{0}}^{\;\;\tau _{1}}.  \label{27x}
\end{eqnarray}

At these point we have all the necessary ingredients (objects and their
properties) for unfolding the irreducible approach. We introduce the
constraints
\begin{equation}
\tilde{\chi}_{\alpha _{0}}=\chi _{\alpha _{0}}+A_{\alpha _{0}}^{\;\;\alpha
_{1}}y_{\alpha _{1}}\approx 0,\;\tilde{\chi}_{\alpha _{2}}=Z_{\;\;\alpha
_{2}}^{\alpha _{1}}y_{\alpha _{1}}\approx 0,  \label{28x}
\end{equation}%
defined on the larger phase-space $\left( z^{\Delta },y_{\alpha _{1}}\right)
$. In the sequel we show that \emph{(\ref{28x}) display all the desired
properties}: equivalence with the intermediate system (\ref{26}),
second-class behaviour, irreducibility, and, most important, the associated
Dirac bracket coincides (weakly) with the original one, corresponding to the
second-order reducible second-class constraints. The proof of all these
properties is contained within the next two theorems.

\begin{theorem}
\label{prop2}Constraints (\ref{28x}) exhibit the following properties:

\noindent (i) equivalence to (\ref{26}), i.e.\footnote{%
Due to the equivalence (\ref{28y}), in what follows we will use the same
symbol of weak equality in relation with each constraint set (\ref{26}) and
respectively (\ref{28x}).}
\begin{equation}
\tilde{\chi}_{\alpha _{0}}\approx 0,\;\tilde{\chi}_{\alpha _{2}}\approx
0\Leftrightarrow \chi _{\alpha _{0}}\approx 0,\;y_{\alpha _{1}}\approx 0;
\label{28y}
\end{equation}

\noindent (ii) second-class behaviour, i.e. the matrix
\begin{equation}
C_{\Delta \Delta ^{\prime }}=\left[ \tilde{\chi}_{\Delta },\tilde{\chi}%
_{\Delta ^{\prime }}\right] ,  \label{28z}
\end{equation}%
is invertible, where
\begin{equation}
\tilde{\chi}_{\Delta }=\left( \tilde{\chi}_{\alpha _{0}},\tilde{\chi}%
_{\alpha _{2}}\right) ;  \label{bv1}
\end{equation}

\noindent (iii) irreducibility.
\end{theorem}

\proof%
(i) It is easy to see that if (\ref{26}) holds, then (\ref{28x}) also holds
\begin{equation}
\chi _{\alpha _{0}}\approx 0,\;y_{\alpha _{1}}\approx 0\Rightarrow \tilde{%
\chi}_{\alpha _{0}}\approx 0,\;\tilde{\chi}_{\alpha _{2}}\approx 0.
\label{wq1}
\end{equation}%
By means of relations (\ref{27y}) and (\ref{27qq}), from (\ref{28x}) we
infer
\begin{equation}
\chi _{\alpha _{0}}=D_{\;\;\alpha _{0}}^{\beta _{0}}\tilde{\chi}_{\beta
_{0}},\;y_{\alpha _{1}}=Z_{\;\;\gamma _{1}}^{\alpha _{0}}\hat{e}_{\;\;\alpha
_{1}}^{\gamma _{1}}\tilde{\chi}_{\alpha _{0}}+A_{\alpha _{1}}^{\;\;\beta
_{2}}\bar{D}_{\;\;\beta _{2}}^{\alpha _{2}}\tilde{\chi}_{\alpha _{2}}.
\label{36}
\end{equation}%
From (\ref{36}) we obtain that if (\ref{28x}) is satisfied, then (\ref{26})
is also valid
\begin{equation}
\tilde{\chi}_{\alpha _{0}}\approx 0,\;\tilde{\chi}_{\alpha _{2}}\approx
0\Rightarrow \chi _{\alpha _{0}}\approx 0,\;y_{\alpha _{1}}\approx 0.
\label{37}
\end{equation}%
Relations (\ref{wq1}) and (\ref{37}) proves (i).

(ii) By means of (\ref{28x}) and (\ref{36}) we find the Poisson brackets
among the functions $\tilde{\chi}_{\Delta }$ in the form
\begin{eqnarray}
\left[ \tilde{\chi}_{\alpha _{0}},\tilde{\chi}_{\beta _{0}}\right] &\approx
&\mu _{\alpha _{0}\beta _{0}}^{\left( 2\right) },\;\left[ \tilde{\chi}%
_{\alpha _{0}},\tilde{\chi}_{\beta _{2}}\right] \approx A_{\alpha
_{0}}^{\;\;\alpha _{1}}\omega _{\alpha _{1}\beta _{1}}Z_{\;\;\beta
_{2}}^{\beta _{1}},  \label{29q} \\
\;\left[ \tilde{\chi}_{\alpha _{2}},\tilde{\chi}_{\beta _{2}}\right]
&\approx &Z_{\;\;\alpha _{2}}^{\alpha _{1}}\omega _{\alpha _{1}\beta
_{1}}Z_{\;\;\beta _{2}}^{\beta _{1}},  \label{29p}
\end{eqnarray}%
where $\mu _{\alpha _{0}\beta _{0}}^{\left( 2\right) }$ is given by (\ref%
{27x}). Then, the matrix $C_{\Delta \Delta ^{\prime }}$ takes the concrete
form
\begin{equation}
C_{\Delta \Delta ^{\prime }}=\left(
\begin{array}{ll}
\mu _{\alpha _{0}\beta _{0}}^{\left( 2\right) } & A_{\alpha
_{0}}^{\;\;\alpha _{1}}\omega _{\alpha _{1}\beta _{1}}Z_{\;\;\beta
_{2}}^{\beta _{1}} \\
Z_{\;\;\alpha _{2}}^{\alpha _{1}}\omega _{\alpha _{1}\beta _{1}}A_{\beta
_{0}}^{\;\;\beta _{1}} & Z_{\;\;\alpha _{2}}^{\alpha _{1}}\omega _{\alpha
_{1}\beta _{1}}Z_{\;\;\beta _{2}}^{\beta _{1}}%
\end{array}%
\right) ,  \label{29z}
\end{equation}%
where $\Delta =\left( \alpha _{0},\alpha _{2}\right) $ indexes the line and $%
\Delta ^{\prime }=\left( \beta _{0},\beta _{2}\right) $ the column. In order
to prove that $C_{\Delta \Delta ^{\prime }}$ is invertible we will simply
exhibit its inverse. Direct computation based on relations (\ref{27qq}), (%
\ref{27qz}), (\ref{27qs}), (\ref{27wp}), and (\ref{27z}) shows that
\begin{equation}
C^{\Delta ^{\prime }\Delta ^{\prime \prime }}=\left(
\begin{array}{ll}
\mu ^{\left( 2\right) \beta _{0}\rho _{0}} & Z_{\;\;\gamma _{1}}^{\beta _{0}}%
\hat{e}_{\;\;\sigma _{1}}^{\gamma _{1}}\omega ^{\sigma _{1}\lambda
_{1}}A_{\lambda _{1}}^{\;\;\tau _{2}}\bar{D}_{\;\;\tau _{2}}^{\rho _{2}} \\
\bar{D}_{\;\;\lambda _{2}}^{\beta _{2}}A_{\sigma _{1}}^{\;\;\lambda
_{2}}\omega ^{\sigma _{1}\lambda _{1}}\hat{e}_{\;\;\lambda _{1}}^{\gamma
_{1}}Z_{\;\;\gamma _{1}}^{\rho _{0}} & \bar{D}_{\;\;\lambda _{2}}^{\beta
_{2}}A_{\sigma _{1}}^{\;\;\lambda _{2}}\omega ^{\sigma _{1}\lambda
_{1}}A_{\lambda _{1}}^{\;\;\tau _{2}}\bar{D}_{\;\;\tau _{2}}^{\rho _{2}}%
\end{array}%
\right) ,  \label{29y}
\end{equation}%
with $\mu ^{\left( 2\right) \beta _{0}\rho _{0}}$ as in (\ref{27x})
satisfies the relations
\begin{equation}
C_{\Delta \Delta ^{\prime }}C^{\Delta ^{\prime }\Delta ^{\prime \prime
}}\approx \left(
\begin{array}{ll}
\delta _{\;\;\alpha _{0}}^{\rho _{0}} & 0 \\
0 & \delta _{\;\;\alpha _{2}}^{\rho _{2}}%
\end{array}%
\right) ,  \label{p11}
\end{equation}%
and hence the matrix of elements (\ref{29z}) is invertible, its inverse
being precisely (\ref{29y}). This proves (ii).

(iii) As the matrix (\ref{29z}) is invertible, it possesses no nontrivial
null vectors. In consequence, the functions $\tilde{\chi}_{\Delta }$ are all
independent, so the constraint set (\ref{28x}) is indeed irreducible. This
proves (iii). $\Box $

By means of result (\ref{29y}), the Dirac bracket associated with the
irreducible second-class constraints (\ref{28x})
\begin{equation}
\left. \left[ F,G\right] ^{\left( 2\right) \ast }\right\vert _{\mathrm{ired}%
}=\left[ F,G\right] -\left[ F,\tilde{\chi}_{\Delta }\right] C^{\Delta \Delta
^{\prime }}\left[ \tilde{\chi}_{\Delta ^{\prime }},G\right] ,  \label{i4}
\end{equation}%
takes the concrete form
\begin{eqnarray}
\left. \left[ F,G\right] ^{\left( 2\right) \ast }\right\vert _{\mathrm{ired}%
} &=&\left[ F,G\right] -\left[ F,\tilde{\chi}_{\alpha _{0}}\right] \mu
^{\left( 2\right) \alpha _{0}\beta _{0}}\left[ \tilde{\chi}_{\beta _{0}},G%
\right] -  \nonumber \\
&&\left[ F,\tilde{\chi}_{\alpha _{0}}\right] Z_{\;\;\gamma _{1}}^{\alpha
_{0}}\hat{e}_{\;\;\sigma _{1}}^{\gamma _{1}}\omega ^{\sigma _{1}\lambda
_{1}}A_{\lambda _{1}}^{\;\;\tau _{2}}\bar{D}_{\;\;\tau _{2}}^{\beta _{2}}%
\left[ \tilde{\chi}_{\beta _{2}},G\right] -  \nonumber \\
&&\left[ F,\tilde{\chi}_{\alpha _{2}}\right] \bar{D}_{\;\;\lambda
_{2}}^{\alpha _{2}}A_{\sigma _{1}}^{\;\;\lambda _{2}}\omega ^{\sigma
_{1}\lambda _{1}}\hat{e}_{\;\;\lambda _{1}}^{\gamma _{1}}Z_{\;\;\gamma
_{1}}^{\beta _{0}}\left[ \tilde{\chi}_{\beta _{0}},G\right] -  \nonumber \\
&&\left[ F,\tilde{\chi}_{\alpha _{2}}\right] \bar{D}_{\;\;\lambda
_{2}}^{\alpha _{2}}A_{\sigma _{1}}^{\;\;\lambda _{2}}\omega ^{\sigma
_{1}\lambda _{1}}A_{\lambda _{1}}^{\;\;\tau _{2}}\bar{D}_{\;\;\tau
_{2}}^{\beta _{2}}\left[ \tilde{\chi}_{\beta _{2}},G\right] .  \label{i5}
\end{eqnarray}%
We observe that the first line from the right-hand side of (\ref{i5}) is
generated by the first-order reducibility relations (see (\ref{31})), while
the remaining terms are due to the second-order reducibility functions.
Together with (\ref{28x}) \emph{formula (\ref{i5}) is the corner stone of
our irreducible approach}. We will show that it coincides (weakly) with the
Dirac bracket of the intermediate system, and therefore with the original
Dirac bracket for the second-order reducible second-class constraints.

\begin{theorem}
\label{dirac}The Dirac bracket with respect to the irreducible second-class
constraints, (\ref{i5}), coincides with that of the intermediate system
\begin{equation}
\left. \left[ F,G\right] ^{\left( 2\right) \ast }\right\vert _{\mathrm{ired}%
}\approx \left. \left[ F,G\right] ^{\left( 2\right) \ast }\right\vert _{z,y}.
\label{32y}
\end{equation}
\end{theorem}

\proof%
In order to prove the theorem we start from the right-hand side of (\ref{i5}%
) and show that it is weakly equal to the right-hand side of (\ref{27}).
Using relations (\ref{27y}), (\ref{27qq}), (\ref{27z}), and (\ref{27x}), by
direct computation we find that
\begin{eqnarray}
\left[ F,\tilde{\chi}_{\alpha _{0}}\right] \mu ^{\left( 2\right) \alpha
_{0}\beta _{0}}\left[ \tilde{\chi}_{\beta _{0}},G\right] &\approx &\left[
F,\chi _{\alpha _{0}}\right] \mu ^{\left( 2\right) \alpha _{0}\beta _{0}}%
\left[ \chi _{\beta _{0}},G\right] +  \nonumber \\
&&\left[ F,y_{\alpha _{1}}\right] D_{\;\;\sigma _{1}}^{\alpha _{1}}\omega
^{\sigma _{1}\lambda _{1}}D_{\;\;\lambda _{1}}^{\beta _{1}}\left[ y_{\beta
_{1}},G\right] ,  \label{32q} \\
\left[ F,\tilde{\chi}_{\alpha _{0}}\right] Z_{\;\;\gamma _{1}}^{\alpha _{0}}%
\hat{e}_{\;\;\sigma _{1}}^{\gamma _{1}}\omega ^{\sigma _{1}\lambda
_{1}}A_{\lambda _{1}}^{\;\;\tau _{2}}\bar{D}_{\;\;\tau _{2}}^{\beta _{2}}%
\left[ \tilde{\chi}_{\beta _{2}},G\right] &\approx &\left[ F,y_{\alpha _{1}}%
\right] D_{\;\;\sigma _{1}}^{\alpha _{1}}\omega ^{\sigma _{1}\lambda
_{1}}\times  \nonumber \\
&&\left( \delta _{\;\;\lambda _{1}}^{\beta _{1}}-D_{\;\;\lambda _{1}}^{\beta
_{1}}\right) \left[ y_{\beta _{1}},G\right] ,  \label{32w} \\
\left[ F,\tilde{\chi}_{\alpha _{2}}\right] \bar{D}_{\;\;\lambda
_{2}}^{\alpha _{2}}A_{\sigma _{1}}^{\;\;\lambda _{2}}\omega ^{\sigma
_{1}\lambda _{1}}\hat{e}_{\;\;\lambda _{1}}^{\gamma _{1}}Z_{\;\;\gamma
_{1}}^{\beta _{0}}\left[ \tilde{\chi}_{\beta _{0}},G\right] &\approx &\left[
F,y_{\alpha _{1}}\right] \left( \delta _{\;\;\sigma _{1}}^{\alpha
_{1}}-D_{\;\;\sigma _{1}}^{\alpha _{1}}\right) \times  \nonumber \\
&&\omega ^{\sigma _{1}\lambda _{1}}D_{\;\;\lambda _{1}}^{\beta _{1}}\left[
y_{\beta _{1}},G\right] ,  \label{32z} \\
\left[ F,\tilde{\chi}_{\alpha _{2}}\right] \bar{D}_{\;\;\lambda
_{2}}^{\alpha _{2}}A_{\sigma _{1}}^{\;\;\lambda _{2}}\omega ^{\sigma
_{1}\lambda _{1}}A_{\lambda _{1}}^{\;\;\tau _{2}}\bar{D}_{\;\;\tau
_{2}}^{\beta _{2}}\left[ \tilde{\chi}_{\beta _{2}},G\right] &\approx &\left[
F,y_{\alpha _{1}}\right] \left( \delta _{\;\;\sigma _{1}}^{\alpha
_{1}}-D_{\;\;\sigma _{1}}^{\alpha _{1}}\right) \times  \nonumber \\
&&\omega ^{\sigma _{1}\lambda _{1}}\left( \delta _{\;\;\lambda _{1}}^{\beta
_{1}}-D_{\;\;\lambda _{1}}^{\beta _{1}}\right) \left[ y_{\beta _{1}},G\right]
.  \label{32x}
\end{eqnarray}%
Inserting the above relations into (\ref{i5}), we find (\ref{32y}). This
proves the theorem. $\Box $

\subsection{Main result}

Combining (\ref{28}) and (\ref{32y}) we reach the result
\begin{equation}
\left[ F,G\right] ^{\left( 2\right) \ast }\approx \left. \left[ F,G\right]
^{\left( 2\right) \ast }\right\vert _{\mathrm{ired}}.  \label{32}
\end{equation}%
\emph{The last formula proves that we can approach second-order reducible
second-class constraints in an irreducible fashion.} Thus, starting with the
second-order reducible constraints (\ref{1}) we construct the irreducible
constraints (\ref{28x}), whose Poisson brackets form an invertible matrix.
Formula (\ref{32}) ensures that the Dirac bracket within the irreducible
setting coincides with that from the reducible version. This is the main
result of the present paper.

Moreover, the new variables, $y_{\alpha _{1}}$, do not affect the
irreducible Dirac bracket as from (\ref{i5}) we have that $\left. \left[
y_{\alpha _{1}},F\right] ^{\left( 2\right) \ast }\right\vert _{\mathrm{ired}%
}\approx 0$. Thus, the equations of motion for the original reducible system
can be written as $\dot{z}^{a}\approx \left. \left[ z^{a},H\right] ^{\left(
2\right) \ast }\right\vert _{\mathrm{ired}}$, where $H$ is the canonical
Hamiltonian. The equations of motion for $y_{\alpha _{1}}$ read as $\dot{y}%
_{\alpha _{1}}\approx 0$, and lead to $y_{\alpha _{1}}=0$ by taking some
appropriate boundary conditions (vacuum to vacuum) for these unphysical
variables. This completes the general procedure.

\section{Example\label{exam}}

We exemplify the general results exposed in the above in the case of a field
theory --- gauge-fixed three-forms, subject to the second-class constraints
\begin{equation}
\chi _{\alpha _{0}}\equiv \left(
\begin{array}{c}
-3\partial ^{i_{3}}\pi _{i_{3}i_{1}i_{2}} \\
-\partial _{j_{3}}A^{j_{3}j_{1}j_{2}}%
\end{array}%
\right) \approx 0.  \label{41x}
\end{equation}%
Thus, the constraints (\ref{41x}) are second-stage reducible, the first-,
respectively, second-stage reducibility matrices being given by
\begin{equation}
Z_{\;\;\alpha _{1}}^{\alpha _{0}}=\left(
\begin{array}{cc}
Z_{k_{1}}^{i_{1}i_{2}} & \mathbf{0} \\
\mathbf{0} & Z_{j_{1}j_{2}}^{l_{1}}%
\end{array}%
\right) ,\;Z_{\;\;\alpha _{2}}^{\alpha _{1}}=\left(
\begin{array}{cc}
Z^{k_{1}} & \mathbf{0} \\
\mathbf{0} & Z_{l_{1}}%
\end{array}%
\right) ,  \label{41y}
\end{equation}%
with
\begin{equation}
Z_{k_{1}}^{i_{1}i_{2}}=\delta _{\;\;k_{1}}^{\left[ i_{1}\right. }\partial
^{\left. i_{2}\right] },\;Z_{j_{1}j_{2}}^{l_{1}}=\delta _{\;\;\left[
j_{1}\right. }^{l_{1}}\partial _{\left. j_{2}\right] },\;Z^{k_{1}}=\partial
^{k_{1}},\;Z_{l_{1}}=\partial _{l_{1}}.  \label{41z}
\end{equation}%
The matrix of the Poisson brackets among the constraints (\ref{41x}) is
expressed by
\begin{equation}
C_{\alpha _{0}\beta _{0}}=\left(
\begin{array}{cc}
\mathbf{0} & \Delta D_{\;\;i_{1}i_{2}}^{i_{3}i_{4}} \\
-\Delta D_{\;\;j_{3}j_{4}}^{j_{1}j_{2}} & \mathbf{0}%
\end{array}%
\right) ,  \label{41w}
\end{equation}%
where
\begin{equation}
D_{\;\;i_{1}i_{2}}^{i_{3}i_{4}}=\frac{1}{2}\left( \delta _{\;\;\left[
i_{1}\right. }^{i_{3}}\delta _{\;\;\left. i_{2}\right] }^{i_{4}}-\frac{%
\delta _{\;\;k}^{\left[ i_{4}\right. }\partial ^{\left. i_{3}\right] }\delta
_{\;\;\left[ i_{2}\right. }^{k}\partial _{\left. i_{1}\right] }}{\Delta }%
\right) ,  \label{41q}
\end{equation}%
and $\Delta =\partial ^{i}\partial _{i}$. If we take
\begin{equation}
A_{\alpha _{1}}^{\;\;\beta _{2}}=\left(
\begin{array}{cc}
Z_{k_{1}} & \mathbf{0} \\
\mathbf{0} & Z^{l_{1}}%
\end{array}%
\right) ,  \label{q41}
\end{equation}%
we obtain
\begin{equation}
D_{\;\;\alpha _{2}}^{\beta _{2}}=Z_{\;\;\alpha _{2}}^{\alpha _{1}}A_{\alpha
_{1}}^{\;\;\beta _{2}}=\left(
\begin{array}{cc}
\Delta & 0 \\
0 & \Delta%
\end{array}%
\right) ,  \label{w41}
\end{equation}%
such that
\begin{equation}
\bar{D}_{\;\;\lambda _{2}}^{\alpha _{2}}=\left(
\begin{array}{cc}
\frac{1}{\Delta } & 0 \\
0 & \frac{1}{\Delta }%
\end{array}%
\right) .  \label{z41}
\end{equation}%
We remark that $A_{\beta _{1}}^{\;\;\beta _{2}}$ given by (\ref{q41}) can be
expressed like in (\ref{27ww}) for
\begin{equation}
\sigma _{\alpha _{1}\beta _{1}}=\left(
\begin{array}{cc}
\mathbf{0} & \delta _{\;\;k_{1}}^{k_{2}} \\
\delta _{\;\;l_{2}}^{l_{1}} & \mathbf{0}%
\end{array}%
\right)  \label{cq1}
\end{equation}%
and
\begin{equation}
\sigma ^{\alpha _{2}\beta _{2}}=\left(
\begin{array}{cc}
0 & 1 \\
1 & 0%
\end{array}%
\right) .  \label{cq2}
\end{equation}%
With the help of (\ref{41y}) and (\ref{q41})--(\ref{z41}), from (\ref{a7})
we find that
\begin{equation}
D_{\;\;\beta _{1}}^{\alpha _{1}}=\left(
\begin{array}{cc}
D_{\;\;k_{2}}^{k_{1}} & \mathbf{0} \\
\mathbf{0} & D_{\;\;l_{1}}^{l_{2}}%
\end{array}%
\right) ,  \label{x41}
\end{equation}%
where
\begin{equation}
D_{\;\;j}^{i}=\delta _{\;\;j}^{i}-\frac{\partial ^{i}\partial _{j}}{\Delta }.
\label{38}
\end{equation}%
On the other hand, we can set $D_{\;\;\beta _{1}}^{\alpha _{1}}$ in the form
expressed by (\ref{1qa}) by choosing
\begin{equation}
\bar{A}_{\beta _{0}}^{\;\alpha _{1}}=\left(
\begin{array}{cc}
\frac{1}{2\Delta }Z_{\;\;i_{3}i_{4}}^{k_{1}} & \mathbf{0} \\
\mathbf{0} & \frac{1}{2\Delta }Z_{\;\;l_{1}}^{j_{3}j_{4}}%
\end{array}%
\right) .  \label{23}
\end{equation}%
Then, it is easy to see that
\begin{equation}
Z_{\;\;\alpha _{1}}^{\alpha _{0}}\bar{A}_{\beta _{0}}^{\;\alpha _{1}}=\left(
\begin{array}{cc}
\frac{1}{2\Delta }\delta _{\;\;k_{1}}^{\left[ i_{2}\right. }\partial
^{\left. i_{1}\right] }\delta _{\;\;\left[ i_{4}\right. }^{k_{1}}\partial
_{\left. i_{3}\right] } & \mathbf{0} \\
\mathbf{0} & \frac{1}{2\Delta }\delta _{\;\;l_{1}}^{\left[ j_{4}\right.
}\partial ^{\left. j_{3}\right] }\delta _{\;\;\left[ j_{2}\right.
}^{l_{1}}\partial _{\left. j_{1}\right] }%
\end{array}%
\right) ,  \label{q23}
\end{equation}%
such that with the aid of (\ref{11f}) we find
\begin{equation}
D_{\;\;\beta _{0}}^{\alpha _{0}}=\left(
\begin{array}{cc}
D_{\;\;i_{3}i_{4}}^{i_{1}i_{2}} & \mathbf{0} \\
\mathbf{0} & D_{\;\;j_{1}j_{2}}^{j_{3}j_{4}}%
\end{array}%
\right) .  \label{w23}
\end{equation}%
Based on the fact that $D_{\;\;i_{3}i_{4}}^{i_{1}i_{2}}$ is a projector,
i.e.
\begin{equation}
D_{\;\;i_{3}i_{4}}^{i_{1}i_{2}}D_{\;\;j_{1}j_{2}}^{i_{3}i_{4}}=D_{\;%
\;j_{1}j_{2}}^{i_{1}i_{2}},  \label{x23}
\end{equation}%
from (\ref{11c}) and (\ref{41w}) we obtain that
\begin{equation}
M^{\left( 2\right) \alpha _{0}\beta _{0}}=\left(
\begin{array}{cc}
\mathbf{0} & -\frac{1}{\Delta }D_{\;\;i_{3}i_{4}}^{i_{1}i_{2}} \\
\frac{1}{\Delta }D_{\;\;j_{1}j_{2}}^{j_{3}j_{4}} & \mathbf{0}%
\end{array}%
\right) .  \label{y23}
\end{equation}%
With the help of (\ref{14q}) and (\ref{y23}) we have that the fundamental
Dirac brackets read as
\begin{equation}
\left[ A^{ijk}\left( x\right) ,\pi _{i^{\prime }j^{\prime }k^{\prime
}}\left( y\right) \right] _{x^{0}=y^{0}}^{\left( 2\right) \ast
}=D_{\;\;i^{\prime }j^{\prime }k^{\prime }}^{ijk}\delta ^{D-1}\left( \mathbf{%
x}-\mathbf{y}\right) ,  \label{v23}
\end{equation}%
\begin{equation}
\left[ A^{ijk}\left( x\right) ,A^{i^{\prime }j^{\prime }k^{\prime }}\left(
y\right) \right] _{x^{0}=y^{0}}^{\left( 2\right) \ast }=0,\;\left[ \pi
_{ijk}\left( x\right) ,\pi _{i^{\prime }j^{\prime }k^{\prime }}\left(
y\right) \right] _{x^{0}=y^{0}}^{\left( 2\right) \ast }=0,  \label{29}
\end{equation}%
where $D_{\;\;i^{\prime }j^{\prime }k^{\prime }}^{ijk}$ is also a projector,
expressed by
\begin{equation}
D_{\;\;i^{\prime }j^{\prime }k^{\prime }}^{ijk}=\frac{1}{3!}\left( \delta
_{\;\;\left[ i^{\prime }\right. }^{i}\delta _{\;\;j^{\prime }}^{j}\delta
_{\;\;\left. k^{\prime }\right] }^{k}-\frac{\partial ^{\left[ i\right.
}\delta _{\;\;l_{1}}^{j}\delta _{\;\;l_{2}}^{\left. k\right] }\partial _{%
\left[ i^{\prime }\right. }\delta _{\;\;j^{\prime }}^{l_{1}}\delta
_{\;\;\left. k^{\prime }\right] }^{l_{2}}}{2\Delta }\right) .  \label{30}
\end{equation}%
Formula (\ref{20}) together with (\ref{w23}) and (\ref{y23}) provides
\begin{equation}
\mu ^{\left( 2\right) \alpha _{0}\beta _{0}}=\left(
\begin{array}{cc}
\mathbf{0} & -\frac{1}{2\Delta }\delta _{\;\;\left[ i_{3}\right.
}^{i_{1}}\delta _{\;\;\left. i_{4}\right] }^{i_{2}} \\
\frac{1}{2\Delta }\delta _{\;\;\left[ j_{1}\right. }^{j_{3}}\delta
_{\;\;\left. j_{2}\right] }^{j_{4}} & \mathbf{0}%
\end{array}%
\right) ,  \label{q30}
\end{equation}%
which clearly exhibits that $\mu ^{\left( 2\right) \alpha _{0}\beta _{0}}$
is invertible. By computing the fundamental Dirac brackets with the help of (%
\ref{24}) (with $\mu ^{\left( 2\right) \alpha _{0}\beta _{0}}$ given by (\ref%
{q30})) we reobtain precisely (\ref{v23})--(\ref{29}).

On the other hand, using the former relation in (\ref{41y}) as well as (\ref%
{y23}) and (\ref{q30}) into (\ref{21z}) produces
\begin{equation}
\tilde{\omega}^{\gamma _{1}\rho _{1}}=\left(
\begin{array}{cc}
\mathbf{0} & \frac{1}{2\Delta ^{2}}\delta _{\;\;m_{2}}^{m_{1}} \\
-\frac{1}{2\Delta ^{2}}\delta _{\;\;n_{1}}^{n_{2}} & \mathbf{0}%
\end{array}%
\right) .  \label{q31}
\end{equation}%
Simple computation shows that $\tilde{\omega}^{\gamma _{1}\rho _{1}}$ given
in (\ref{q31}) is in agreement with (\ref{27w}) if we take
\begin{equation}
\hat{e}_{\;\;\sigma _{1}}^{\gamma _{1}}=\left(
\begin{array}{cc}
-\frac{1}{2\Delta }\delta _{\;\;p_{1}}^{m_{1}} & \mathbf{0} \\
\mathbf{0} & -\frac{1}{\Delta }\delta _{\;\;n_{1}}^{s_{1}}%
\end{array}%
\right)  \label{q32}
\end{equation}%
and
\begin{equation}
\omega ^{\sigma _{1}\tau _{1}}=\left(
\begin{array}{cc}
\mathbf{0} & \delta _{\;\;p_{2}}^{p_{1}} \\
-\delta _{\;\;s_{1}}^{s_{2}} & \mathbf{0}%
\end{array}%
\right) .  \label{q33}
\end{equation}%
Consequently, the inverse of $\hat{e}_{\;\;\sigma _{1}}^{\gamma _{1}}$ of
the form (\ref{q32}) reads as
\begin{equation}
\hat{E}_{\;\;\tau _{1}}^{\sigma _{1}}=\left(
\begin{array}{cc}
-2\delta _{\;\;p_{2}}^{p_{1}}\Delta & \mathbf{0} \\
\mathbf{0} & -\delta _{\;\;s_{1}}^{s_{2}}\Delta%
\end{array}%
\right) .  \label{q34}
\end{equation}%
Using (\ref{x41}), (\ref{q32}), and (\ref{q34}) we deduce that relation (\ref%
{27qq}) is automatically verified. Based on formula (\ref{27y}), from (\ref%
{23}) and (\ref{q34}) it follows that
\begin{equation}
A_{\alpha _{0}}^{\;\;\alpha _{1}}=\left(
\begin{array}{cc}
-Z_{\;\;i_{1}i_{2}}^{k_{1}} & \mathbf{0} \\
\mathbf{0} & -\frac{1}{2}Z_{l_{1}}^{j_{1}j_{2}}%
\end{array}%
\right) .  \label{43}
\end{equation}%
We remark that $A_{\alpha _{0}}^{\;\;\alpha _{1}}$ from (\ref{43}) is
expressed like in (\ref{27qw}) for $\sigma ^{\alpha _{1}\beta _{1}}$ taken
as the inverse of (\ref{cq1}) and
\begin{equation}
\sigma _{\alpha _{0}\beta _{0}}=\left(
\begin{array}{cc}
\mathbf{0} & -\frac{1}{2}\delta _{\;\;\left[ i_{1}\right. }^{i_{3}}\delta
_{\;\;\left. i_{2}\right] }^{i_{4}} \\
-\frac{1}{4}\delta _{\;\;\left[ j_{3}\right. }^{j_{1}}\delta _{\;\;\left.
j_{4}\right] }^{j_{2}} & \mathbf{0}%
\end{array}%
\right) .  \label{q51}
\end{equation}%
The variables $y_{\alpha _{1}}$ in the case of the model under investigation
are given by
\begin{equation}
y_{\alpha _{1}}=\left(
\begin{array}{l}
\pi _{k_{1}} \\
A^{l_{1}}%
\end{array}%
\right) ,  \label{54}
\end{equation}%
where $A^{k}$ is a vector field and $\pi _{k}$ its momentum, conjugated in
the Poisson bracket induced by (\ref{q33}). Replacing (\ref{41x}), (\ref{43}%
), and (\ref{54}) in the first relation from (\ref{28x}), we find the
concrete form of the irreducible constraints $\tilde{\chi}_{\alpha
_{0}}\approx 0$
\begin{eqnarray}
\tilde{\chi}_{i_{1}i_{2}}^{(1)} &\equiv &-3\partial ^{i_{3}}\pi
_{i_{3}i_{1}i_{2}}-\partial _{\left[ i_{1}\right. }\pi _{\left. i_{2}\right]
}\approx 0,  \label{58} \\
\tilde{\chi}^{(2)j_{1}j_{2}} &\equiv &-\partial _{j_{3}}A^{j_{3}j_{1}j_{2}}-%
\frac{1}{2}\partial ^{\left[ j_{1}\right. }A^{\left. j_{2}\right] }\approx 0.
\label{59}
\end{eqnarray}%
Substituting the second relation from (\ref{41y}) together with (\ref{54})
in the second formula from (\ref{28x}) we find the irreducible constraints $%
\tilde{\chi}_{\alpha _{2}}\approx 0$ for the model under study as
\begin{equation}
\tilde{\chi}^{(1)}\equiv \partial ^{k_{1}}\pi _{k_{1}}\approx 0,\;\tilde{\chi%
}^{(2)}\equiv \partial _{l_{1}}A^{l_{1}}\approx 0.  \label{72}
\end{equation}%
At this stage we have constructed all the objects entering the structure of
the irreducible Dirac bracket (\ref{i5}). \emph{It is essential to remark
that the irreducible second-class constraints are local.} If we construct
the irreducible Dirac bracket and evaluate the fundamental Dirac brackets
among the original variables, then we finally obtain that these are
expressed by relations (\ref{v23})--(\ref{29}). This completes the analysis
of gauge-fixed three-form gauge fields.

\section{Conclusion\label{conc}}

To conclude with, in this paper we have exposed an irreducible procedure for
approaching systems with second-order reducible second-class constraints.
Our strategy includes three main steps. First, we express the Dirac bracket
for the reducible system in terms of an invertible matrix. Second, we
establish the equality between this Dirac bracket and that corresponding to
the intermediate theory, based on the constraints (\ref{26}). Third, we
prove that there exists an irreducible second-class constraint set
equivalent with (\ref{26}) such that the corresponding Dirac brackets
coincide. These three steps enforce the fact that the fundamental Dirac
brackets with respect to the original variables derived within the
irreducible and original reducible settings coincide. Moreover, the newly
added variables do not affect the Dirac bracket, so the canonical approach
to the initial reducible system can be developed in terms of the Dirac
bracket corresponding to the irreducible theory. The general procedure was
exemplified on gauge-fixed three-forms. Our procedure does not spoil other
important symmetries of the original system, such as spacetime locality for
second-class field theories.

\section*{Acknowledgment}

This work has been supported in part by the contract
2-CEx-06-11-92/19.09.2006 with the Romanian Ministry of Education and
Research (M.Ed.C.) and by the European Commission FP6 program
MRTN-CT-2004-005104.

\end{document}